\documentclass[%
 reprint,
 amsmath,amssymb,
 aps,
prb,
]{revtex4-2}

\usepackage{amsmath}
\usepackage{mathtools}
\usepackage{xcolor}
\usepackage{graphicx}
\usepackage{dcolumn}
\usepackage{bm}
\usepackage{float}

\begin{document}

\preprint{APS/123-QED}
\title{Going Beyond the Cumulant Approximation II:\\Power Series Correction to Single Particle Green's Function in 1D Holstein Chain}
\author{Bipul Pandey}
\email{bipulpandey2004@uchicago.edu}
\author{Peter B. Littlewood}
\affiliation{James Franck Institute and Department of Physics, University of Chicago, Chicago, IL, 60637, USA}

\date{\today}
\begin{abstract}
Previously \cite{pandey_going_2022}, we introduced a method for systematically correcting a quasiparticle green's function via a power series expansion. Here we present an ODE based formalisms of power series correction that goes beyond the cumulant approximation and implement it to 1D Holstein chain for a wide range of coupling strengths in a scalable and inexpensive fashion at both zero and finite temperature. We show that this first differential formalism of the power series is both qualitatively and quantitatively in excellent agreement with exact diagonalization results on 1D Holstein chain with dispersive bosons for a large range of electron-boson coupling strength. We investigate carrier mass growth rate and carrier energy displacement across a wide range of coupling strength. Finally, we present a heuristic argument which predicts most of the rich satellite structure without explicit calculation.
\end{abstract}

\maketitle


\section{ Introduction}

Electrons in a solid suffer dynamical scattering from collective modes (e.g. phonons, magnons, and plasmons)\cite{ma_formation_2020, dally_three-magnon_2020,lemell_real-time_2015} that are often overlooked in ab-initio modeling of materials. And in many cases there are good reasons for it. Often times, electronic part of the problem takes the front seat in governing macroscopic material properties. The collective excitation of the system which are bosonic in nature merely result in small corrections to the carrier properties. Hence, they take a back seat regardless of their origin (plasma oscillation, spin waves, molecular stretching etc). In such systems, it is not only justifiable but also pragmatic to take into account only the electronic part of the problem in modeling and predicting material properties such as effective mass of the carrier, energy levels, lifetimes etc. This is because of the associated overhead cost of incorporating bosonic degrees of freedom which are extremely expensive but ultimately insignificant in the actual system and predictions thereof. For this reason, development of efficient and accurate methods to handle electronic problem has been a topic of great interest for more than half a century. This has led to the development of myriads of methods from families of Density functional based methods \cite{perdew_density_1985, gonze_adiabatic_1995} to families of Green's function based methods such as GW \cite{deslippe_berkeleygw_2012, govoni_large_2015} that have proven their worth in computing the charged electronic excitation spectra and and predicting properties across a wide variety of materials. 

Nonetheless, in experiments, the collective modes indeed renormalize the carrier(electron or hole) particle's energy, effective mass and lifetime transmuting this particle into a quasiparticle. Another hallmark of carrier-boson coupling is the presence of satellite shake-offs in the electronic spectra\cite{guzzo_valence_2011}. These satellites drain spectral weight from the quasiparticle and somewhat resemble the quasiparticle band structure. Hence they are also termed replica bands. At weak carrier-boson coupling when the bosonic corrections to the carrier are small, non-self consistent post processing methods called cumulant expansions \cite{gunnarsson_corrections_1994,kas_cumulant_2014,zhou_cumulant_2018,caruso_band_2015} have been developed that wrap around GW methods and produce bosonic satellites at roughly the right energy. In recent years, this has proven successful in incorporating plasmonic as well as phonon effects in the electronic band structure. However, cumulant methods are ad hoc by construction and it has been shown that attempts to incorporate higher order corrections or introduce self-consistency in cumulant based methods result in divergence or artifacts \cite{robinson_cumulant_2022,robinson_cumulant_2022-1} of the correction function. On the other hand, going the route of Monte Carlo approach \cite{bonca_spectral_2019} or any basis dependent method like exact diagonalization \cite{barisic_calculation_2004}, coherent basis expansion \cite{de_filippis_static_2005}, although stable, is extremely expensive and thus unscalable to large systems. 

There are physical systems where the coupling is not small. Recent advances in organic (polymeric) semiconductors such as poly(3-hexylthiophene) (P3HT) \cite{pingel_comprehensive_2013, ghosh_spectral_2018, chen_experimental_2021, hulea_tunable_2006} show that, in these long chain systems with vibrational modes (stretching) encompassing a huge numbers of sites, bosonic effects on carrier are significant for two reasons. Firstly, these bosonic stretching modes themselves are highly energetic in nature with energies (~0.2 eV in P3HT) comparable to electronic energy scale (~0.6 - 1 eV in P3HT). Secondly, these modes are strongly coupled to the electronic degrees of freedom \cite{ghosh_excitons_2020}. This has also been observed in proteins \cite{acbas_optical_2014,ing_going_2018,hekstra_electric-field-stimulated_2016} where the electronic degrees of freedom and the collective modes are intimately intertwined. Strong electron phonon coupling is also observed in correlated metals such as cuprates \cite{, damascelli_angle-resolved_2003,baldini_electronphonon-driven_2020, cataudella_temperature_2007}, the iron pnictide high temperature superconductors \cite{rettig_electronphonon_2013,das_small_1993}, the colossal magneto-resistance manganites, and nickelates \cite{millis_double_1995, millis_dynamic_1996, zaanen_freezing_1994, kaplan_mechanism_1998}. Sometimes the electron-phonon coupling conspires with strong correlation physics to produce metal to insulator or metal to bad metal phase transitions. Therefore focusing solely on the electronic problem in these systems is myopic at best and catastrophic at worst because of strong modulation of carrier's energy levels, mobility, masses, and lifetimes by bosons.

On the engineering front, fine control over doping has allowed for the advent of plasmonic devices \cite{tang_plasmonic_2020, cushing_progress_2016,ma_energy_2016} where where plasmons (plasma oscillations) and their coupling to carrier is leveraged for tasks ranging from sensing to energy capture and harvest. Another recent advent is the field of phonovoltaics \cite{melnick_phonovoltaic_2016, melnick_phonovoltaic_2016-2} where the carriers excited after gulping modulated phonon quanta are collected and used for energy generation. At extremely values of carrier-boson coupling, the carrier becomes strongly self trapped and localized creating polaronic states with properties very different than free carriers\cite{riley_crossover_2018, ma_formation_2020}.

The 1D Holstein chain model \cite{holstein_studies_1959} provides a simple yet suitable proving ground to build and benchmark a method to effectively capture the effect of bosons on carrier (electron or hole) band structure. In this model, a single electron is allowed to hop between nearest neighbor sites in a chain of vibrating atomic sites while being coupled to these vibrations (vibrons or phonons). Previously, we developed a power series based method that accurately captured the Green's and the spectral function on a Holstein dimer \cite{pandey_going_2022}. In this work, we generalize our method to a macroscopic 1D Holstein chain. Since the 1D Holstein Hamiltonian also serves as the model that well describes polymeric semiconductors \cite{ghosh_excitons_2020}, the results we show, especially at strong coupling, is more than an exercise in method building and serves as a stepping stone to build intuition and perform practical calculations in these polymeric systems. Moreover, our method is agnostic of the details of non-interacting electronic and bosonic parts of the system and therefore extendable to models beyond the Holstein chain.
 
The paper is organized as follows. In section II, we discuss the Holstein model and introduce the concepts of electron and phonon Green's function, spectral function and electron self energy. We also introduce the Power series correction and the coupling strength. In section III, we briefly introduce the integral formalism of power series and build the first differential formalism from it. Then in section IV and V, we discuss the zero and finite temperature results and explore the dependence of electronic spectral properties on the coupling constant and temperature. In section VI, we validate the differential power series method against exact diagonalization for intermediate coupling strength and show that our method indeed captures the overall spectral shape as well as most of the larger features. In VII, we discuss our method at extreme coupling strengths. Finally in VIII, we provide a heuristic argument to build intuitions about predicting spectral features without doing explicit calculations.

\section{Introduction to the problem} 
The 1D Holstein Hamiltonian has three distinct pieces - the electronic part, the bosonic part, and the electron-boson interaction;
\begin{equation}
    H = H_e + H_{bos} + H_{e-bos}
\end{equation}
The electronic part is given by a simple tight binding model with an electron living in a 1D Bravais lattice of spacing `a' with N sites. Each lattice site is a potential well that localizes the electron on the site with energy $\epsilon_o$ - the on-site energy. The particles can also hop between nearest neighbor wells with hopping energy $-t_{el}$. With electronic ladder operators $c_i^\dagger/c_i$ at site `i', the electronic Hamiltonian in site basis (real space) for this model is;
\begin{equation}
    H_e = \epsilon_o\sum_{i}c_i^\dagger c_i -t_{el} \sum_{i}(c_i^\dagger c_{i+1} + c_{i+1}^\dagger c_i)
    \label{electron}
\end{equation}
The electronic system is submerged in a bath of N boson species with ladder operators $b_j^\dagger/b_j$ which can be thought of as phonons (or vibrons) associated to the vibrational modes of the N sites. The phonons have a bare phonon energy of $\tilde{\omega}$. We also have phonon-phonon interaction with energy $t_{d}$ which introduces phonon dispersion giving rise to a non-trivial phonon band structure similar to real materials. This bosonic Hamiltonian is also of tight binding form.
\begin{equation}
H_{bos} = \tilde{\omega} \sum_{j}b_j^\dagger b_j + {t_{d}}\sum_{j}(b_j^\dagger b_{j+1} + b_{j+1}^\dagger b_j)
    \label{boson}
\end{equation}
The final ingredient is the electron-boson interaction piece with coupling constant g which modulates the on-site electronic energy wherever the electron is present by boson creation and annihilation at that site.
\begin{equation}
    H_{e-bos} = g\sum_{i} c_i^\dagger c_i (b_i^\dagger + b_i)
    \label{electron-boson}
\end{equation}
 In the momenta space, these Hamiltonian pieces \eqref{electron} , \eqref{boson}, \eqref{electron-boson} translate into the following:
 \begin{eqnarray}
 \begin{aligned}
 &H = H_{e} + H_{bos} + H_{e-bos} \quad \text{with}\\
  &H_e = \sum_k \varepsilon_k c_k^\dagger c_k \quad \text{with,}\,\,
    \varepsilon_k = \epsilon_o - 2t_{el}\, cos(ka)\\
    &H_{bos} = \sum_{q}\omega_{q} b_q^\dagger b_q \quad \text{where,}\,\, \omega_q = \tilde{\omega} + 2t_{d} cos(qa)\\
    &H_{e-bos} = \frac{g}{\sqrt{N}} \sum_{k,q}c_{k-q}^\dagger c_k(b_q^\dagger + b_{-q}) 
\end{aligned}
\label{momenta space hamiltonian}
 \end{eqnarray}
 
 As discussed in \cite{kas_cumulant_2014, guzzo_valence_2011, pandey_going_2022, robinson_cumulant_2022}, we will use the retarded Green's function formalism. The quantity of interest are the finite temperature (T) electron and the boson green's functions ($G$ and $\mathcal{D}$ respectively)for a single electron system with the thermal trace over states with zero electron but unrestricted boson number. At zero temperature, the only state that survives in this trace is the electron-boson Fock vacuum $|0\rangle$. We will denote this thermal average by angle brackets with inverse temperature $\beta$ in the subscript ($`\langle\,\,\rangle_{\beta}'$) \cite{mahan_many-particle_2000}.
 \begin{equation}
\begin{aligned}
    G(k,t) &= -i\theta(t)\frac{Tr[e^{-\beta H} \{c_k(t),c_k^{\dagger}\}]}{Tr(e^{-\beta H})}\\
            &=-i\theta(t)\langle\{c_k(t) ,c_k^{\dagger}\}\rangle_{\beta}\\
    \mathcal{D}(q,t) &= -i \theta(t)\langle [A_q (t) ,A_{-q}]\rangle_{\beta} \quad\text{where,}\\ 
     A_q(t) &= (b_q e^{-i\omega_q t} + b_{-q}^\dagger e^{i\omega_q t})
\end{aligned}
\label{Greens definition}
\end{equation}
The non interacting (g=0) electron and the boson Green's function for a single electron system with finite temperature(T) and the Bose occupation factor $N^T_q = (e^{\omega_q/{k_b T}}-1)^{-1}$ are defined as follows\cite{mahan_many-particle_2000}: 
\begin{eqnarray}
\begin{aligned}
G_o(k,t) &= -i \theta(t) e^{-i\varepsilon_k t}\\
\mathcal{D}_o(q,t) &= -i\theta(t)\big[(N^T_q +1)e^{-i\omega_q|t|} + N^T_q e^{i\omega_q|t|}\big]\\
\label{bare Greens}
\end{aligned}
\end{eqnarray}
The first term in $\mathcal{D}_o$ with prefactor $(N_q^T +1)$ model stokes scattering processes which are possible at any temperature while the second term with prefactor $N_q^T$ model anti-stokes processes which are possible only at finite temperature \cite{luchner_phonon_1979}. The spectral function $A(k,\omega)$, which is the quantity measured in experiments is defined following our previous convention \cite{pandey_going_2022} as,
 \begin{equation}
     A(k,\omega)= \frac{1}{\pi} |Im \,G(k,\omega)|
     \label{Spectral Funciton}
 \end{equation}
 
At zero coupling (q=0), there is no electron-phonon interaction in the system and hence the electronic state has infinite lifetime with energy $\varepsilon_k$. However, once the interaction is turned on, the electron-phonon coupling can kick the electron out of its current state into any other state energetically available in the momenta space. Furthermore, depending on the coupling strength, a proportional correction to the pure electronic state's energy and lifetime occurs. This transmutes the infinitely long lived electron/holes into quasi electron/holes with finite lifetime. In this method, we assume that the non-interacting electron Green's function ($G_o$) smoothly transforms to the interacting Green's function ($G$) through a general power series $\mathcal{P}_k$ of the coupling constant squared ($\frac{g^2}{N}$).
\begin{equation}
    G(k,t) = G_o(k,t)\sum_{n=0}^\infty \big(\frac{g^{2}}{N}\big)^n C_n(k,t) = G_o(k,t) \mathcal{P}_k(t)
\label{Power Series definition}
\end{equation}
In this expansion, $C_0=1$ and all other $C_{n\neq0}$ are assumed to be independent functions that vanish at $t=0$. For a given momentum k, the associated power series correction inherits the following temporal contraction relation from the interacting and non-interacting Green's function.
\begin{equation}
    \mathcal{P}_k(t_2-t_1) = \mathcal{P}_k(t_2-t) \mathcal{P}_k(t-t_1) \quad; t_1\leq t \leq t_2
\label{temporal contraction}
\end{equation}
This relation is valid only for power series pieces of same momenta and is essential in producing the cumulant diagrams. All the information about the possible interactions that can change the energy as well as the shorten the lifetime of pure electronic states is packaged together in electron self energy. This electron self energy also has an electron Green's function inside it by definition and hence it too gets a power series correction piece as shown below. With the Green's functions \eqref{Greens definition},the first order self energy $\Sigma(k,t)$ is defined as;
\begin{equation}
\begin{aligned}
    -i\Sigma(k,t) &= \frac{g^2}{N}\sum_{q}\mathcal{D}_o(q,t)G(k-q,t)\\
                &= \frac{g^2}{N}\sum_{q}\mathcal{D}_o(q,t)G_o(k-q,t)\mathcal{P}_{k-q}(t)\\
     &= -i\frac{g^2}{N}\sum_{q}\Sigma_{k,q}(t)\mathcal{P}_{k-q}(t)
\end{aligned}
\end{equation}
The Dyson's equation governs the evolution of the electron green's function by repeated application of self energy to itself.
\begin{equation}
    G(k,\omega) = G_o(k,\omega) + G_o(k,\omega)\Sigma(k,\omega)G(k,\omega)
    \label{Dyson}
\end{equation}
In cumulant approximation, we assume $\mathcal{P}_k = e^C_k$, expand and truncate Dyson's equation in time domain to first order and compute $C_k$. This truncation as well its ad hoc nature makes the cumulant unsystematic as well as unsuitable to handle strong coupling cases.

There are three energy scales in this problem - the electronic, the bosonic and the coupling scale. The electronic energy is captured by the band width ($2t_{el}$). The bosonic energy scale is captured by the average boson energy (geometric average of extreme $\omega_q$ in \eqref{momenta space hamiltonian} to capture boson dispersion). And the coupling scale is captured by the coupling constant `g'. The coupling strength ($\lambda$) captures how big the coupling scale is compared to electronic and the bosonic scales and is the metric of importance for electron-boson interaction in such systems\cite{bonca_spectral_2019, marchand_effect_2013, robinson_cumulant_2022}.
\begin{equation}
\begin{aligned}
\lambda &= \frac{\text{Coupling scale}}{\text{electronic scale}} \times \frac{\text{Coupling scale}}{\text{bosonic scale}}\\
    &= \frac{g}{2t_{el}}\times \frac{g}{\sqrt{\tilde{\omega}^2 - (2t_{d}^2)}}
\end{aligned}
    \label{Coupling strength}
\end{equation}
\begin{figure*}[ht]
    \centering
    \includegraphics[height = 9 cm,width = 1\textwidth]{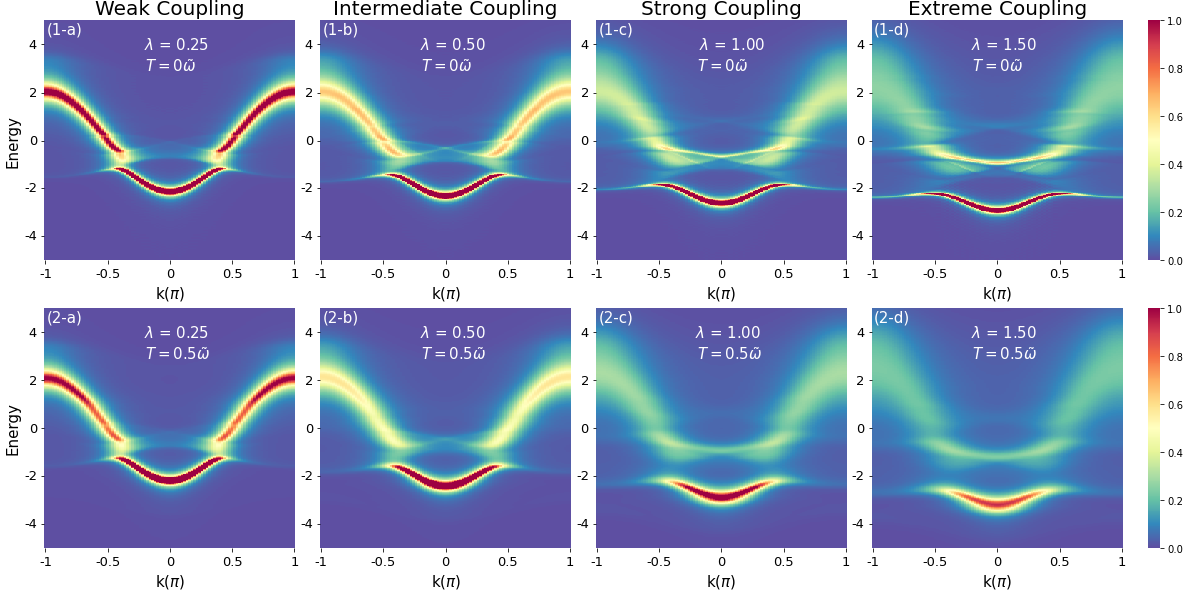}
    \caption{Spectral Function for 100 site single electron 1D Holstein chain. The coupling strength goes from $\lambda = 0.25$ (weak) to $\lambda = 1.5$ (extreme) from left to right column. The temperature goes in the units of bare boson frequency($\Tilde{\omega}$) from $T=0\,\Tilde{\omega}$ to $T=0.5\,\Tilde{\omega}$ from top to bottom row.}
    \label{Spectra_table}
\end{figure*}

\section{Integral and differential forms of Power Series Correction}
In this section we briefly outline the general power series method and direct the reader's attention to our previous work \cite{pandey_going_2022} for more details on the derivation of the power series integral equation as well as discussion on the drawbacks of popular methods. Following the same logic as \cite{pandey_going_2022}, we derive the integral power series equation \eqref{Integral Formalism} for this problem. The self correction term $P^{SC}_k$ incorporates the correction due to the the $q=0$ bosons that do not cause an electronic state change after electron-boson interaction. The effect of these bosons is like that of bosons in the core-hole problem where diminishing replicas of the electronic peak are produced at the boson frequency. The inter-orbital correction term $P^{IC}_k$ incorporates the correction due to all other $q\neq 0$ bosons which change the electronic state during interaction. 
\begin{equation}
\begin{aligned}
    \mathcal{P}_k(t) &= 1+  P^{SC}_k(t) + \sum_{q\neq 0} P^{IC}_k(q,t)\\
    P^{SC}_k(t)&= -i\frac{g^2}{N}\int\displaylimits_{0}^{t}\!\!\! dt_2\!\!\int\displaylimits_{0}^{t_2}\!\!\! d\tau e^{i\varepsilon_k \tau}\Sigma_{k,q=0}(\tau)\mathcal{P}_k(t_2)\\
    P^{IC}_k(q,t) &=-i\frac{g^2}{N}\! \int\displaylimits_{0}^{t}\!\!\!dt_2\!\! \int\displaylimits_{0}^{t_2}\!\!\! d\tau e^{i\varepsilon_k\tau}\Sigma_{k,q}(\tau)\mathcal{P}_{k-q}(\tau)\mathcal{P}_k(t_2-\tau)\\
\label{Integral Formalism}
\end{aligned}
\end{equation}

Previously, we had a simple dimer \cite{pandey_going_2022} and thus it was easy to self consistently solve this integral problem with two coupled corrections starting with the initial guess of $\mathcal{P}_k(t) =1$ for all time. However, this recipe is difficult to implement when we have N coupled functions that we need to self-consistently compute for the entire time domain. This is especially relevant when the coupling strength is not small and the correction functions are wildly fluctuating before they stabilize to their equilibrium values. For large N and $\lambda$, the integral method often gets stuck between oscillating families of solution or diverges for coarse time steps. In this work, we surpass this limitation by constructing differential formulations of the power series. The first differential formalism of power series correction is obtained by applying the fundamental theorem of calculus to \eqref{Integral Formalism}. 
\begin{eqnarray}
\begin{aligned}
\frac{d\mathcal{P}_k(t)}{dt} &= \Big(\frac{-ig^2}{N}\Big)\bigg[\int\displaylimits_{0}^{t}d\tau e^{i\varepsilon_k \tau}\Sigma_{k,q=0}(\tau)\bigg]\mathcal{P}_k(t) \quad+ \\
 \Big(\frac{-ig^2}{N}\Big)&\sum_{q\neq0}\int\displaylimits_{0}^{t}d\tau e^{i\varepsilon_k \tau}\Sigma_{k,q}(\tau)\mathcal{P}_{k-q}(\tau)\mathcal{P}_k(t-\tau)\\
\mathcal{P}_k(t=&0) = 1 \quad \quad \quad\text{(Initial condition)}
\end{aligned}
\label{Differential Formalism}
\end{eqnarray}

The initial condition above is set by letting $t=0$ in the definition of Power series correction \eqref{Power Series definition}. This is consistent with our assumption that the mapping of $G_o$ to $G$ is adiabatic and smooth. The integro-differential equation thus obtained is called a continuous delay differential equation because the value of the correction function at a time-instant depends on its past values - here the $\mathcal{P}_k(t-\tau)$ at the end of \eqref{Differential Formalism}.   

 So far, the equations \eqref{Integral Formalism} and \eqref{Differential Formalism} are formally equivalent. But, \eqref{Differential Formalism} provides the advantage of computing $\mathcal{P}_k$ for each time step gradually (method of step) by recycling its previous values for the delayed coupling rather than having to optimize for an appropriate value of $\mathcal{P}_k$ for the entire time domain at once. We cannot reduce this equation further without severe approximations to the delay term. Assuming that the delay term $\mathcal{P}_k(t-\tau)$ is splittable into $\mathcal{P}_k(t)/\mathcal{P}_k(\tau)$ using relation \eqref{temporal contraction}, we can further simplify the power series into a second differentiable formalism which is much closer in spirit with the self consistent cumulant formalism \cite{robinson_cumulant_2022}. But due to the severity of this approximation, this second order formalism is marred with sudden onset of divergence in the correction function. Further discussion therefore is relegated to appendix \ref{Second differential formalism}.
 
 Once the power series correction is found, the corrected temporal Green's function $G(k,t)$ is calculated by using \eqref{Power Series definition} and Fourier transformed to obtain the frequency space Green's function $G(k,\omega)$ as well as the spectral function \eqref{Spectral Funciton}.

\section{Electron Green's function in Holstein Chain at Zero Temperature}
\begin{figure}[ht]
    \centering
    \includegraphics[height =8 cm, width = 0.48\textwidth]{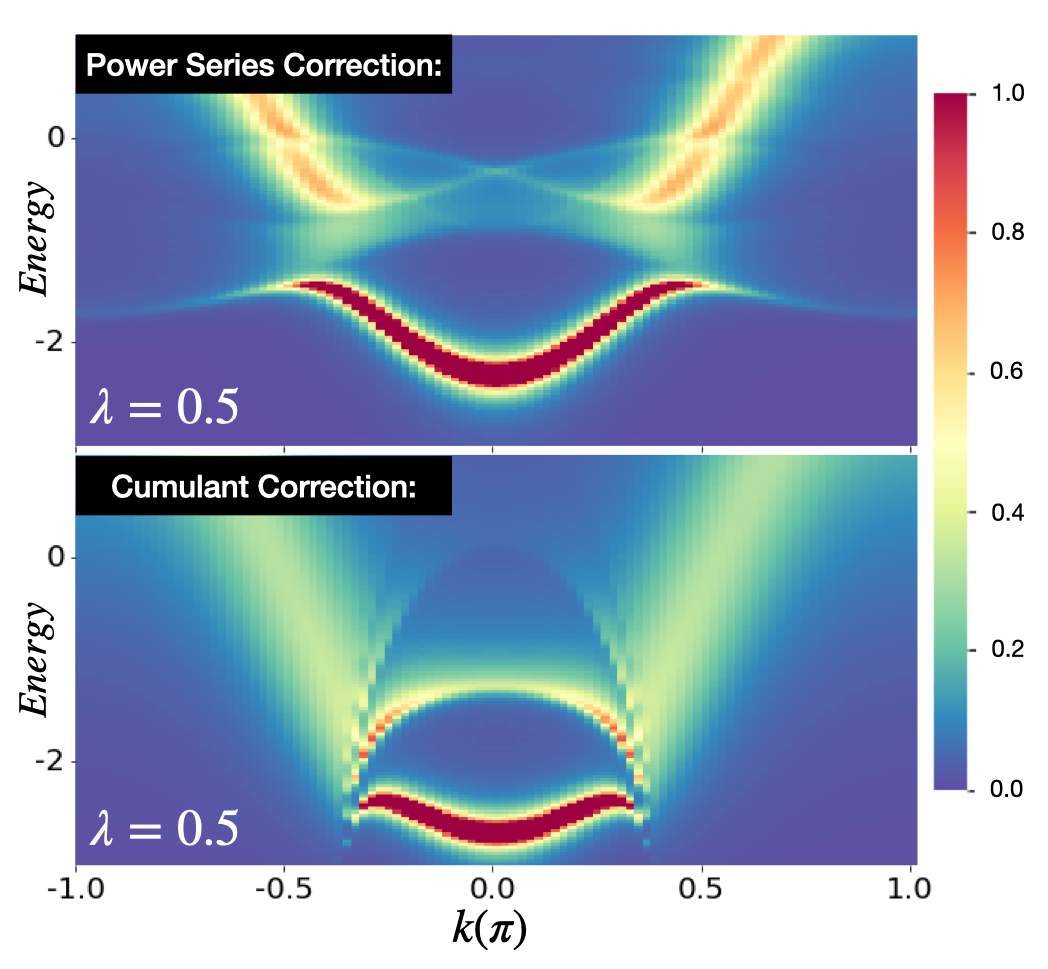}
    \caption{ The head and the first satellite structure as produced by power series and cumulant expansion for $\lambda =0.5$, $\epsilon_o = 0 ,t_el = 1, \tilde{\omega} = 1$ and $t_d = 0.2$. The curvature and the extent of the head is very different for the two approximations. The power series satellite also shows a richer structure with multiple crossing and variation in spectral weight as compared to cumulant.}
    \label{Richer satellite structure}
\end{figure}
The zero temperature Holstein Dimer case for a wide range of coupling  strength was successfully handled by the integral power series in our previous work \cite{pandey_going_2022}. In figure \ref{Spectra_table} first row (1-a to 1-d), we show the implementation of the single differential formalism of the power series on a 1-D Holstein chain with N = 100 sites at zero temperature for different electron-boson coupling strengths ranging from weak ($\lambda =0.25$) to extreme and unrealistic ($\lambda = 1.5$). At weak coupling, the non-interacting sinusoidal electronic band fractures roughly at bare phonon energy $\Tilde{\omega}$ from the band bottom separating the head from the two arms on each side. 
 
\subsection{\label{richer structure}Richer Structure in the Satellites}
Due to boson-mediated electronic excitations spectral weight diffuses from the fermion band to form faint satellites above the head with intricate  crossing structures due to the non-interacting fermion band folding along fracture at the single boson excitation level on both sides as seen in figure \ref{Richer satellite structure}. The cumulant expansion, in the end, is the very first order approximation of the power series \cite{pandey_going_2022}. And hence, it is only able to produce an averaged structure unlike the full Power series. Therefore, it is not surprising that cumulant only puts an averaged lump to represent a satellite with a much richer structures and spread. Furthermore, the extent of the lower head is also severely restricted by the cumulant and the intensity of the fractured arms are also significantly lower in cumulant corrected band structure. 
\subsection{Effect of coupling strength on carrier properties}
 \begin{figure}[ht]
     \centering
     \includegraphics[height = 9 cm,width = 0.48\textwidth]{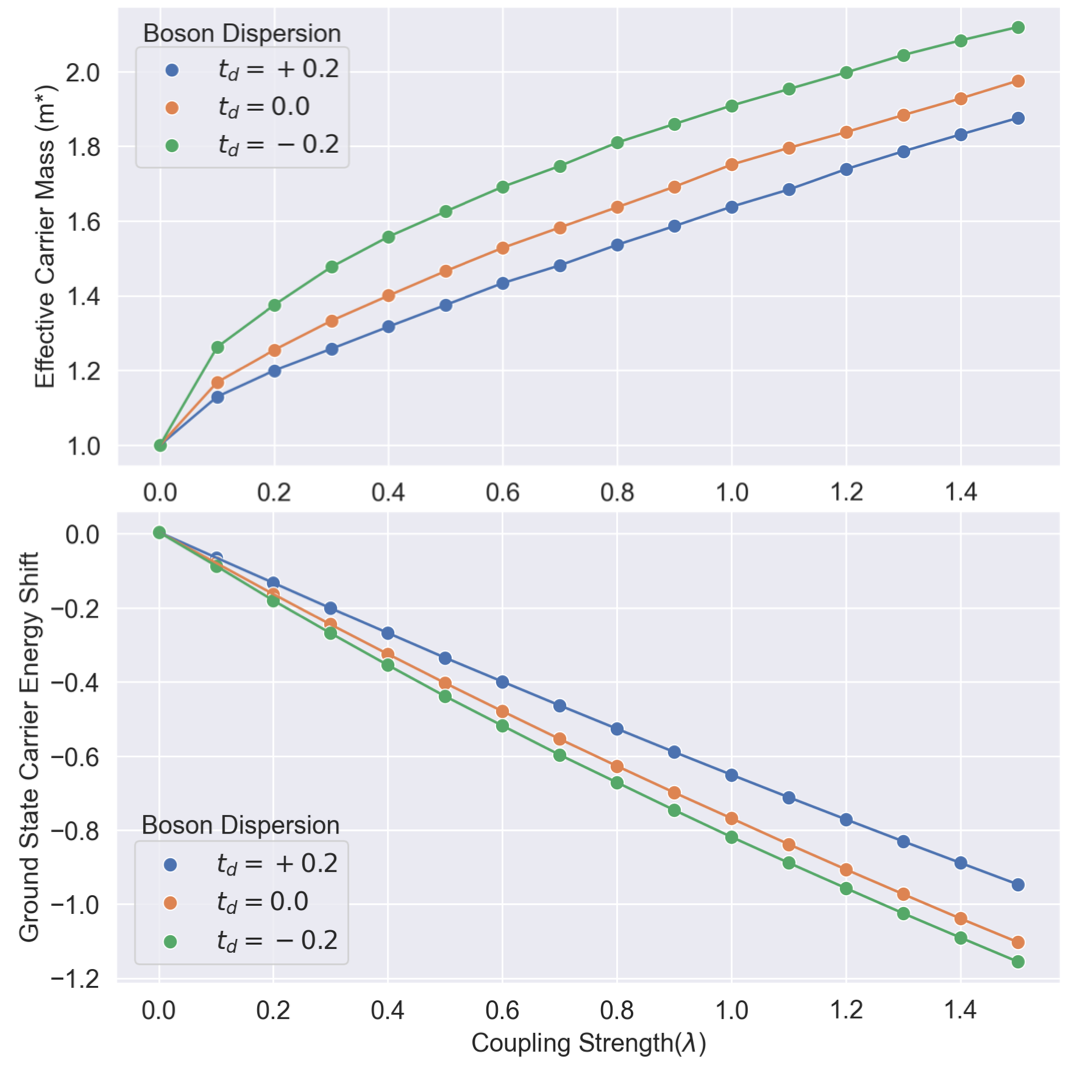}
     \caption{Effective carrier mass (top) shows initially grows rapidly but quickly becomes linear with  coupling strength $\lambda$. The ground state carrier energy (band bottom) also gets displaced linearly to lower energy as a function ($\lambda$). The rates of change of both properties depend on boson dispersion($t_d$).}
     \label{Carrier Properties}
 \end{figure}

 The coupling to boson also significantly affects the electronic band's curvature as well as the location of the band bottom. As seen in figure \ref{Spectra_table}, stronger electron-boson coupling ($\lambda$) proportionally flattens and pushes the fractured head of the electronic band away from the non-interacting parabolic fermionic band.  The carrier's effective mass is given by the inverse of the band curvature. This effective mass initially grows rapidly but quickly becomes linear with growing coupling strength $\lambda$ . The head of the band which gives the carrier's ground state energy linearly increases with the coupling strength. That being said, there are differences in the effective mass growth rate as well as the band displacement rate depending on the boson dispersion `$t_d$' as seen in figure \ref{Carrier Properties}. A larger carrier effective mass and a smaller carrier energy displacement is observed for smaller $|\omega_{q=0}|$ mode in \eqref{momenta space hamiltonian} because of the single boson excitation level (and thus band splitting) being closer to the non-interacting band bottom.
\section{The Effect of Temperature on Electron Green's function}
For finite temperature, the main advantage of our method is that the onus of carrying the thermal information lies on the non-interacting Green's functions rather than the power series machinery. Hence, the computation speed and resources required to calculate the finite temperature interacting Green's function remain more-or-less same as that in zero temperature interacting Green's function. This is far from the case of Monte Carlo methods \cite{bonca_dynamic_2021} or exact diagonalization because of the thermal trace over infinitely many states in \eqref{Greens definition} due to finite temperature. Here we present the interacting Green's function for finite temperature ($T= 0.5 \tilde{\omega}$) at different coupling regimes for N=100 site Holstein chain as shown in the second row of figure \ref{Spectra_table}. The main effect of temperature is to broaden and stretch the overall band structure on both sides of the fracture because of the onset of finite temperature anti-stokes scattering processes.
\section{Validation of Power Series correction}
In order to validate our results, we rely upon exact diagonalization results in the same system. The problem with exact diagonalization is its steeply rising cost with each new site and/or each additional boson in the system. This is because of the quadratic dependence of the Hamiltonian matrix on both these factors. Hence, we perform exact diagonalization on 8 site Holstein chain in a finite boson basis. Convergence of features is estimated by increasing the number of bosons. For intermediate coupling of $\lambda = 0.5$ and 3 bosons for each momenta value $q$, the majority of the features converge and the differences remain only in the smaller features. 
\begin{figure}[ht]
    \centering
    \includegraphics[height = 6 cm,width = 0.48\textwidth]{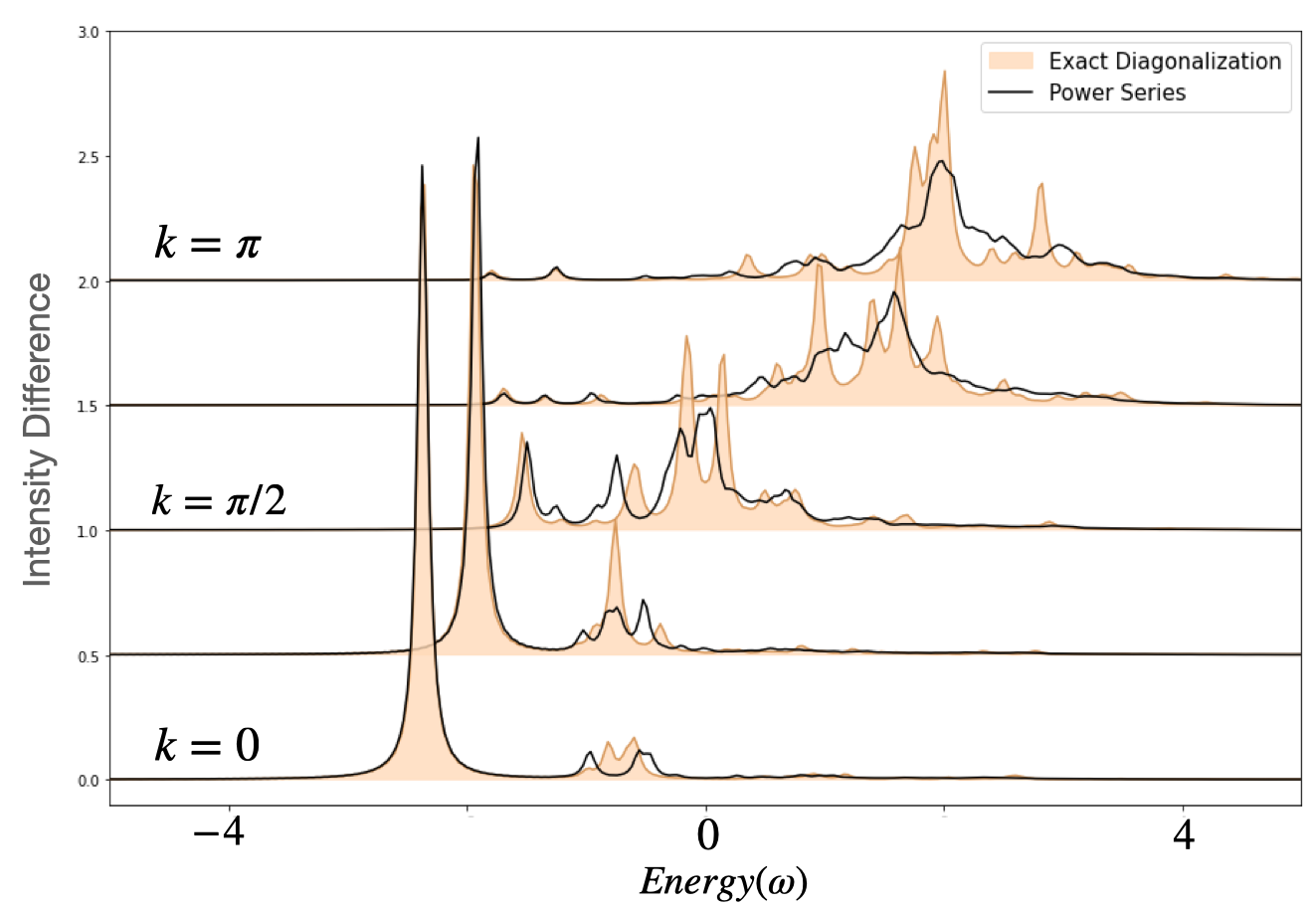}
    \caption{Power series (line) closely matches most of the exact diagonalization features (solid) for 8 site Holstein chain at $\lambda = 0.5$.}
    \label{Validation}
\end{figure}
As seen in figure \ref{Validation}, Power series captures the overall shape of the spectra as well as most of the larger details of the exact diagonalization. This is specially true of the head of the spectra where power series almost matched exact diagonalization. Here, power series performs this calculation within few minutes while the exact diagonalization takes numerous hours to complete.

\section{Electron Green's function at extreme coupling strengths}
With increasing coupling strength, the system adiabatically moves to a region where Migdal's theorem \cite{migdal_interaction_1958,gunnarsson_corrections_1994} becomes invalid. This is because by approximating the self energy with only the very first irreducible self energy diagram, we neglect many higher order irreducible diagrams onward of second order. These diagrams at higher coupling have a non-negligible contribution to particle dynamics. Hence, because of their absence, the spectral function computed with the first order self energy is deficient with missing features, incorrect feature sizes and incorrect separations between features. This also is the reason why exact diagonalization is not feasible in this regime because extremely high number of bosons are necessary for the validity of the calculated spectra.

Nevertheless, using power series correction with first order self energy we can assess general features and trends the actual spectra should follow at extreme coupling ($\lambda = 1-1.5$) for zero or finite temperature as shown in figure \ref{Spectra_table}. It is surprising that at extreme values of coupling, rather than being more intricate, the satellite structure coagulates and simplifies to look more like a `replica' of the fermion band's head. Apart from this, there is a significant damping of the fractured arms and a significant increase in the spectral weight of this replica and a flattening of the band head. Of course these results must be taken with a grain of salt because of the aforementioned reasons regarding the validity of Migdal's approximation. But this exercise proves that differential power series method is inherently stable even at extreme values of coupling. Therefore given a more exhaustive set of self energy diagrams, power series machinery can always produce a better electron spectral function.

\section {Heuristic Argument for origin of structures}
\begin{figure}
    \centering
    \includegraphics[height = 5.3 cm,width = 0.48\textwidth]{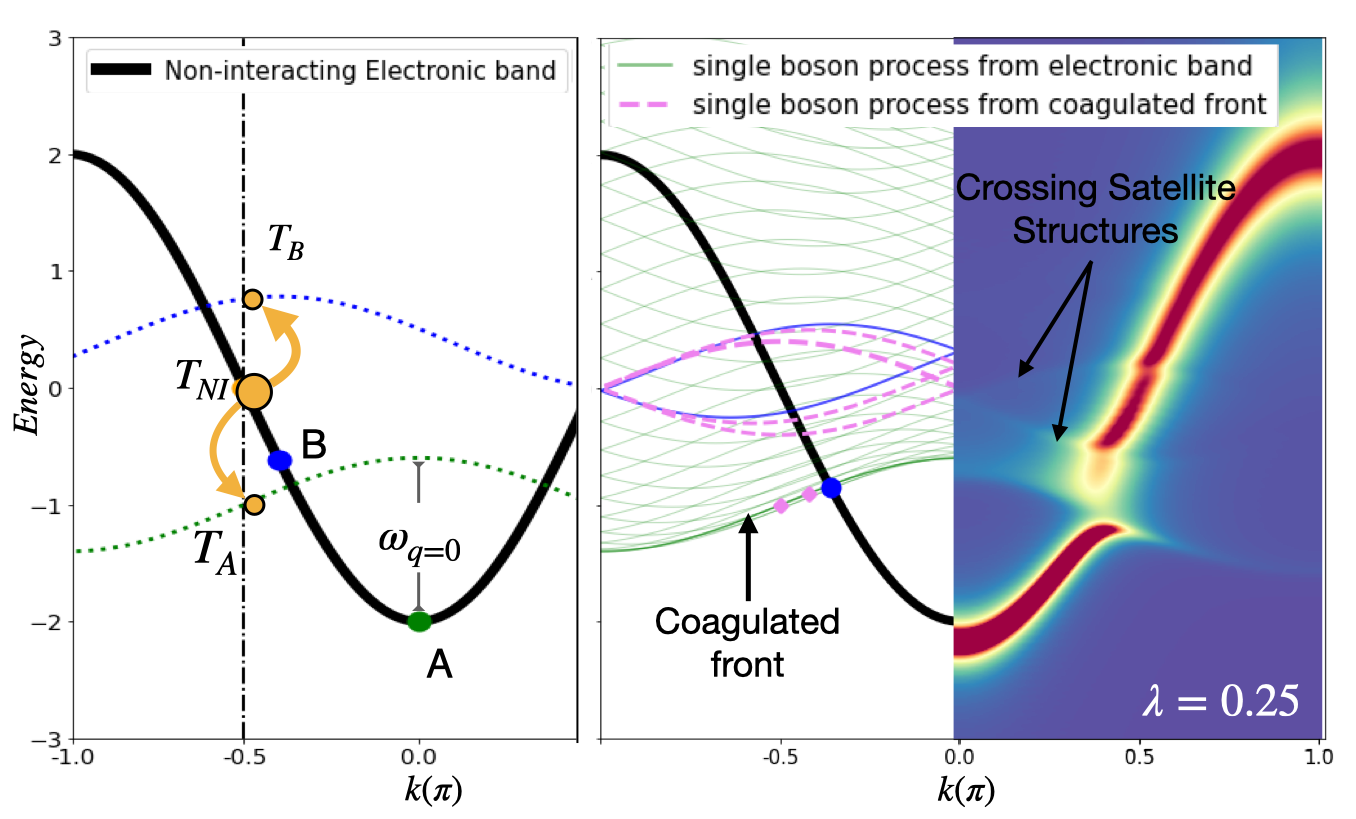}
    \caption{(On left) The spectral weight from non-interacting (NI) target point $T_{NI}$ flows to $T_A$ and $T_B$ because of the presence of points A and B respectively in the non-interacting band structure. The dotted lines are boson band structure centered at A and B at a distance $\omega_{q=0}$ away. (On right) Boson band structure drawn over uniform sampling of non-interacting electronic band reveals coagulated front of spectral weights from single boson process from NI band. The NI band fractures and bends around this coagulated front. Single boson process from this front form the crossing satellite structures.}
    \label{heuristic construction}
\end{figure}
We claim that we can predict most of the locations where spectral weight from any non-interacting $\varepsilon_k$ flows towards as well as identify regions where novel structures (fractures , crosses in satellite structures) occur by simply collecting the complex exponential terms from \eqref{Differential Formalism} and \eqref{Greens definition}.
\begin{equation*}
    \text{Structures over $\varepsilon_k$}\approx \sum_{q}(-ie^{-i\varepsilon_k t})\big[e^{i\varepsilon_k t} e^{-i (\varepsilon_{k-q} + \omega_{k-q}) t}\big]
\end{equation*}
The first term $(-ie^{-i\varepsilon^{k}t})$ is just the non-interacting band structure at momenta k. The second term here represents the erasure some of spectral weight from the non-interacting band at k. The third term $e^{-i (\varepsilon_{k-q} + \omega_{k-q}) t}$ represents the effect of the rest of the band structure at momenta $k$ where the aforementioned erased spectral weight flows to. This term is equivalent to drawing a boson band structure centered at every point on the non-interacting band at a distance $\omega_{q=0}$ away as shown in figure \ref{heuristic construction}. 

If we uniformly sample the non-interacting(NI) electronic band structure and follow the recipe above, we see that the spectral weights from a single boson event originating from the non-interacting band coagulates to form a front shown in \ref{heuristic construction}. At zero temperature, the electronic band structure is pristine below this front and heavily modified above this front. The electronic band fractures and folds along the creases formed by this coagulated front in our calculation. Single boson events originating from this front creates the rich structure of satellite as discussed in \ref{richer structure}.   

Through this work, we showed that differential power series provides an efficient, scalable and inexpensive way to incorporate electron-boson physics into non-interacting electronic band structure. This method is agnostic of the details of electronic and bosonic physics. Furthermore, application of power series correction on 1D Holstein model for a wide range of coupling strength shows that this method is stable. Power series was also validated against exact diagonalization where power series qualitatively and quantitatively resolved most of the spectral features.

The dependence of effective mass on coupling strength as well as boson dispersion for 1D Holstein chain was studied through power series spectral function. We also constructed a heuristic argument to quickly access interacting electronic band structure given non-interacting electronic and bosonic band structures. Finally, This was used to explain the rich spectral structure of Holstein polaron problem in 1D. Future direction includes using the power series machinery to study electronic properties of organic semiconductors as well as extending this machinery beyond Holstein model to incorporate a broader class of problems. 
\appendix
\section{Second differential formalism and issues with stability}
\label{Second differential formalism}
We can make one more approximation to the \eqref{Differential Formalism} by inserting the temporal contraction property \eqref{temporal contraction} to simplify the delay term $\mathcal{P}_k(t-\tau)$ as $\mathcal{P}_k(t)/\mathcal{P}_k(\tau)$ in equation \eqref{Differential Formalism}. This simplification allows for the use of fundamental theorem of calculus again to generate a second differential formalism of power series correction. Inspired from the cumulant approximation where the cumulant correction derivative is zero at $t=0$, we get a second initial condition where the power series derivative is zero at $t=0$ .
\begin{eqnarray}
\begin{aligned}
&\frac{d^2}{dt^2}ln[\mathcal{P}_k(t)] = \Big(\frac{-ig^2}{N}\Big) \sum_{\forall q} e^{i\varepsilon_k t}\Sigma_{k,q}(t) \,\,\frac{\mathcal{P}_{k-q}(t)}{\mathcal{P}_k(t)}\\
&\mathcal{P}_k(t=0) =1 \quad \text{and,}\quad \left.\frac{d\mathcal{P}_k(t)}{dt}\right\vert_{t=0} =0
\end{aligned}
\label{Double Differential}
\end{eqnarray}

We can abridge this further by setting the correction to by cumulant like i.e $\mathcal{P}_k(t)= e^{C_k(t)}$.
\begin{eqnarray}
\begin{aligned}
&\frac{d^2C_k(t)}{dt^2} = \Big(\frac{-ig^2}{N}\Big) \sum_{\forall q} e^{i\varepsilon_k t}\Sigma_{k,q}(t) \,\,e^{C_{k-q}(t) - C_k(t)}\\
&C_k(t=0) = 0 \quad\text{and,}\quad \left.\frac{dC_k(t)}{dt}\right\vert_{t=0} = 0
\end{aligned}
\label{Self consistent cumulant}
\end{eqnarray}

The expressions \eqref{Double Differential}, \eqref{Self consistent cumulant} are closer in spirit to the expression derived by \cite{robinson_cumulant_2022} in their theory of self-consistent cumulants in that in both expressions, the delay term has been severely modified by assuming that the temporal contraction relation \eqref{temporal contraction} can be inverted. However, during numerical implementation, both \eqref{Double Differential} and \eqref{Self consistent cumulant} as well as the self consistent cumulant formalism \cite{robinson_cumulant_2022} are susceptible to sudden onset of divergence in solution. This is because of the correction term on right term $\mathcal{P}_k(t)$ inversely relating to the second derivative of its own logarithm on the left in equation \eqref{Double Differential}. Therefore a slight increase in the correction term from its equilibrium value amplifies the its second derivative and the correction soon diverges. This is seen as we take smaller and smaller grid sizes because the divergence also gets pushed further in time.

\nocite{*}

\bibliography{Bib_for_Differential_Power_Series_Paper}

\begin{thebibliography}{47}%
\makeatletter
\providecommand \@ifxundefined [1]{%
 \@ifx{#1\undefined}
}%
\providecommand \@ifnum [1]{%
 \ifnum #1\expandafter \@firstoftwo
 \else \expandafter \@secondoftwo
 \fi
}%
\providecommand \@ifx [1]{%
 \ifx #1\expandafter \@firstoftwo
 \else \expandafter \@secondoftwo
 \fi
}%
\providecommand \natexlab [1]{#1}%
\providecommand \enquote  [1]{``#1''}%
\providecommand \bibnamefont  [1]{#1}%
\providecommand \bibfnamefont [1]{#1}%
\providecommand \citenamefont [1]{#1}%
\providecommand \href@noop [0]{\@secondoftwo}%
\providecommand \href [0]{\begingroup \@sanitize@url \@href}%
\providecommand \@href[1]{\@@startlink{#1}\@@href}%
\providecommand \@@href[1]{\endgroup#1\@@endlink}%
\providecommand \@sanitize@url [0]{\catcode `\\12\catcode `\$12\catcode
  `\&12\catcode `\#12\catcode `\^12\catcode `\_12\catcode `\%12\relax}%
\providecommand \@@startlink[1]{}%
\providecommand \@@endlink[0]{}%
\providecommand \url  [0]{\begingroup\@sanitize@url \@url }%
\providecommand \@url [1]{\endgroup\@href {#1}{\urlprefix }}%
\providecommand \urlprefix  [0]{URL }%
\providecommand \Eprint [0]{\href }%
\providecommand \doibase [0]{https://doi.org/}%
\providecommand \selectlanguage [0]{\@gobble}%
\providecommand \bibinfo  [0]{\@secondoftwo}%
\providecommand \bibfield  [0]{\@secondoftwo}%
\providecommand \translation [1]{[#1]}%
\providecommand \BibitemOpen [0]{}%
\providecommand \bibitemStop [0]{}%
\providecommand \bibitemNoStop [0]{.\EOS\space}%
\providecommand \EOS [0]{\spacefactor3000\relax}%
\providecommand \BibitemShut  [1]{\csname bibitem#1\endcsname}%
\let\auto@bib@innerbib\@empty
\bibitem [{\citenamefont {Pandey}\ and\ \citenamefont
  {Littlewood}(2022)}]{pandey_going_2022}%
  \BibitemOpen
  \bibfield  {author} {\bibinfo {author} {\bibfnamefont {B.}~\bibnamefont
  {Pandey}}\ and\ \bibinfo {author} {\bibfnamefont {P.~B.}\ \bibnamefont
  {Littlewood}},\ }\href {https://doi.org/10.1103/PhysRevLett.129.136401}
  {\bibfield  {journal} {\bibinfo  {journal} {Physical Review Letters}\
  }\textbf {\bibinfo {volume} {129}},\ \bibinfo {pages} {136401} (\bibinfo
  {year} {2022})}\BibitemShut {NoStop}%
\bibitem [{\citenamefont {Ma}\ \emph {et~al.}(2020)\citenamefont {Ma},
  \citenamefont {Cheng}, \citenamefont {Tian}, \citenamefont {Liu},
  \citenamefont {Cui}, \citenamefont {Huang}, \citenamefont {Tan},
  \citenamefont {Yang},\ and\ \citenamefont {Wang}}]{ma_formation_2020}%
  \BibitemOpen
  \bibfield  {author} {\bibinfo {author} {\bibfnamefont {X.}~\bibnamefont
  {Ma}}, \bibinfo {author} {\bibfnamefont {Z.}~\bibnamefont {Cheng}}, \bibinfo
  {author} {\bibfnamefont {M.}~\bibnamefont {Tian}}, \bibinfo {author}
  {\bibfnamefont {X.}~\bibnamefont {Liu}}, \bibinfo {author} {\bibfnamefont
  {X.}~\bibnamefont {Cui}}, \bibinfo {author} {\bibfnamefont {Y.}~\bibnamefont
  {Huang}}, \bibinfo {author} {\bibfnamefont {S.}~\bibnamefont {Tan}}, \bibinfo
  {author} {\bibfnamefont {J.}~\bibnamefont {Yang}},\ and\ \bibinfo {author}
  {\bibfnamefont {B.}~\bibnamefont {Wang}},\ }\bibfield  {journal} {\bibinfo
  {journal} {Nano Letters}\ }\href
  {https://doi.org/10.1021/acs.nanolett.0c03802} {10.1021/acs.nanolett.0c03802}
  (\bibinfo {year} {2020}),\ \bibinfo {note} {publisher: American Chemical
  Society}\BibitemShut {NoStop}%
\bibitem [{\citenamefont {Dally}\ \emph {et~al.}(2020)\citenamefont {Dally},
  \citenamefont {Heng}, \citenamefont {Keselman}, \citenamefont {Bordelon},
  \citenamefont {Stone}, \citenamefont {Balents},\ and\ \citenamefont
  {Wilson}}]{dally_three-magnon_2020}%
  \BibitemOpen
  \bibfield  {author} {\bibinfo {author} {\bibfnamefont {R.~L.}\ \bibnamefont
  {Dally}}, \bibinfo {author} {\bibfnamefont {A.~J.}\ \bibnamefont {Heng}},
  \bibinfo {author} {\bibfnamefont {A.}~\bibnamefont {Keselman}}, \bibinfo
  {author} {\bibfnamefont {M.~M.}\ \bibnamefont {Bordelon}}, \bibinfo {author}
  {\bibfnamefont {M.~B.}\ \bibnamefont {Stone}}, \bibinfo {author}
  {\bibfnamefont {L.}~\bibnamefont {Balents}},\ and\ \bibinfo {author}
  {\bibfnamefont {S.~D.}\ \bibnamefont {Wilson}},\ }\href
  {https://doi.org/10.1103/PhysRevLett.124.197203} {\bibfield  {journal}
  {\bibinfo  {journal} {Physical Review Letters}\ }\textbf {\bibinfo {volume}
  {124}},\ \bibinfo {pages} {197203} (\bibinfo {year} {2020})},\ \bibinfo
  {note} {publisher: American Physical Society}\BibitemShut {NoStop}%
\bibitem [{\citenamefont {Lemell}\ \emph {et~al.}(2015)\citenamefont {Lemell},
  \citenamefont {Neppl}, \citenamefont {Wachter}, \citenamefont {Tőkési},
  \citenamefont {Ernstorfer}, \citenamefont {Feulner}, \citenamefont
  {Kienberger},\ and\ \citenamefont {Burgdörfer}}]{lemell_real-time_2015}%
  \BibitemOpen
  \bibfield  {author} {\bibinfo {author} {\bibfnamefont {C.}~\bibnamefont
  {Lemell}}, \bibinfo {author} {\bibfnamefont {S.}~\bibnamefont {Neppl}},
  \bibinfo {author} {\bibfnamefont {G.}~\bibnamefont {Wachter}}, \bibinfo
  {author} {\bibfnamefont {K.}~\bibnamefont {Tőkési}}, \bibinfo {author}
  {\bibfnamefont {R.}~\bibnamefont {Ernstorfer}}, \bibinfo {author}
  {\bibfnamefont {P.}~\bibnamefont {Feulner}}, \bibinfo {author} {\bibfnamefont
  {R.}~\bibnamefont {Kienberger}},\ and\ \bibinfo {author} {\bibfnamefont
  {J.}~\bibnamefont {Burgdörfer}},\ }\href
  {https://doi.org/10.1103/PhysRevB.91.241101} {\bibfield  {journal} {\bibinfo
  {journal} {Physical Review B}\ }\textbf {\bibinfo {volume} {91}},\ \bibinfo
  {pages} {241101} (\bibinfo {year} {2015})},\ \bibinfo {note} {publisher:
  American Physical Society}\BibitemShut {NoStop}%
\bibitem [{\citenamefont {Perdew}(1985)}]{perdew_density_1985}%
  \BibitemOpen
  \bibfield  {author} {\bibinfo {author} {\bibfnamefont {J.~P.}\ \bibnamefont
  {Perdew}},\ }\bibfield  {journal} {\bibinfo  {journal} {International Journal
  of Quantum Chemistry}\ }\textbf {\bibinfo {volume} {28}},\ \href
  {https://doi.org/10.1002/qua.560280846} {10.1002/qua.560280846} (\bibinfo
  {year} {1985})\BibitemShut {NoStop}%
\bibitem [{\citenamefont {Gonze}(1995)}]{gonze_adiabatic_1995}%
  \BibitemOpen
  \bibfield  {author} {\bibinfo {author} {\bibfnamefont {X.}~\bibnamefont
  {Gonze}},\ }\href {https://doi.org/10.1103/PhysRevA.52.1096} {\bibfield
  {journal} {\bibinfo  {journal} {Physical Review A}\ }\textbf {\bibinfo
  {volume} {52}},\ \bibinfo {pages} {1096} (\bibinfo {year} {1995})},\ \bibinfo
  {note} {publisher: American Physical Society}\BibitemShut {NoStop}%
\bibitem [{\citenamefont {Deslippe}\ \emph {et~al.}(2012)\citenamefont
  {Deslippe}, \citenamefont {Samsonidze}, \citenamefont {Strubbe},
  \citenamefont {Jain}, \citenamefont {Cohen},\ and\ \citenamefont
  {Louie}}]{deslippe_berkeleygw_2012}%
  \BibitemOpen
  \bibfield  {author} {\bibinfo {author} {\bibfnamefont {J.}~\bibnamefont
  {Deslippe}}, \bibinfo {author} {\bibfnamefont {G.}~\bibnamefont
  {Samsonidze}}, \bibinfo {author} {\bibfnamefont {D.~A.}\ \bibnamefont
  {Strubbe}}, \bibinfo {author} {\bibfnamefont {M.}~\bibnamefont {Jain}},
  \bibinfo {author} {\bibfnamefont {M.~L.}\ \bibnamefont {Cohen}},\ and\
  \bibinfo {author} {\bibfnamefont {S.~G.}\ \bibnamefont {Louie}},\ }\href
  {https://doi.org/https://doi.org/10.1016/j.cpc.2011.12.006} {\bibfield
  {journal} {\bibinfo  {journal} {Computer Physics Communications}\ }\textbf
  {\bibinfo {volume} {183}},\ \bibinfo {pages} {1269} (\bibinfo {year}
  {2012})}\BibitemShut {NoStop}%
\bibitem [{\citenamefont {Govoni}\ and\ \citenamefont
  {Galli}(2015)}]{govoni_large_2015}%
  \BibitemOpen
  \bibfield  {author} {\bibinfo {author} {\bibfnamefont {M.}~\bibnamefont
  {Govoni}}\ and\ \bibinfo {author} {\bibfnamefont {G.}~\bibnamefont {Galli}},\
  }\href {https://doi.org/10.1021/ct500958p} {\bibfield  {journal} {\bibinfo
  {journal} {Journal of Chemical Theory and Computation}\ }\textbf {\bibinfo
  {volume} {11}},\ \bibinfo {pages} {2680} (\bibinfo {year} {2015})},\ \bibinfo
  {note} {publisher: American Chemical Society}\BibitemShut {NoStop}%
\bibitem [{\citenamefont {Guzzo}\ \emph {et~al.}(2011)\citenamefont {Guzzo},
  \citenamefont {Lani}, \citenamefont {Sottile}, \citenamefont {Romaniello},
  \citenamefont {Gatti}, \citenamefont {Kas}, \citenamefont {Rehr},
  \citenamefont {Silly}, \citenamefont {Sirotti},\ and\ \citenamefont
  {Reining}}]{guzzo_valence_2011}%
  \BibitemOpen
  \bibfield  {author} {\bibinfo {author} {\bibfnamefont {M.}~\bibnamefont
  {Guzzo}}, \bibinfo {author} {\bibfnamefont {G.}~\bibnamefont {Lani}},
  \bibinfo {author} {\bibfnamefont {F.}~\bibnamefont {Sottile}}, \bibinfo
  {author} {\bibfnamefont {P.}~\bibnamefont {Romaniello}}, \bibinfo {author}
  {\bibfnamefont {M.}~\bibnamefont {Gatti}}, \bibinfo {author} {\bibfnamefont
  {J.~J.}\ \bibnamefont {Kas}}, \bibinfo {author} {\bibfnamefont {J.~J.}\
  \bibnamefont {Rehr}}, \bibinfo {author} {\bibfnamefont {M.~G.}\ \bibnamefont
  {Silly}}, \bibinfo {author} {\bibfnamefont {F.}~\bibnamefont {Sirotti}},\
  and\ \bibinfo {author} {\bibfnamefont {L.}~\bibnamefont {Reining}},\ }\href
  {https://doi.org/10.1103/PhysRevLett.107.166401} {\bibfield  {journal}
  {\bibinfo  {journal} {Physical Review Letters}\ }\textbf {\bibinfo {volume}
  {107}},\ \bibinfo {pages} {166401} (\bibinfo {year} {2011})},\ \bibinfo
  {note} {publisher: American Physical Society}\BibitemShut {NoStop}%
\bibitem [{\citenamefont {Gunnarsson}\ \emph {et~al.}(1994)\citenamefont
  {Gunnarsson}, \citenamefont {Meden},\ and\ \citenamefont
  {Schönhammer}}]{gunnarsson_corrections_1994}%
  \BibitemOpen
  \bibfield  {author} {\bibinfo {author} {\bibfnamefont {O.}~\bibnamefont
  {Gunnarsson}}, \bibinfo {author} {\bibfnamefont {V.}~\bibnamefont {Meden}},\
  and\ \bibinfo {author} {\bibfnamefont {K.}~\bibnamefont {Schönhammer}},\
  }\href {https://doi.org/10.1103/PhysRevB.50.10462} {\bibfield  {journal}
  {\bibinfo  {journal} {Physical Review B}\ }\textbf {\bibinfo {volume} {50}},\
  \bibinfo {pages} {10462} (\bibinfo {year} {1994})}\BibitemShut {NoStop}%
\bibitem [{\citenamefont {Kas}\ \emph {et~al.}(2014)\citenamefont {Kas},
  \citenamefont {Rehr},\ and\ \citenamefont {Reining}}]{kas_cumulant_2014}%
  \BibitemOpen
  \bibfield  {author} {\bibinfo {author} {\bibfnamefont {J.~J.}\ \bibnamefont
  {Kas}}, \bibinfo {author} {\bibfnamefont {J.~J.}\ \bibnamefont {Rehr}},\ and\
  \bibinfo {author} {\bibfnamefont {L.}~\bibnamefont {Reining}},\ }\href
  {https://doi.org/10.1103/PhysRevB.90.085112} {\bibfield  {journal} {\bibinfo
  {journal} {Physical Review B}\ }\textbf {\bibinfo {volume} {90}},\ \bibinfo
  {pages} {085112} (\bibinfo {year} {2014})}\BibitemShut {NoStop}%
\bibitem [{\citenamefont {Zhou}\ \emph {et~al.}(2018)\citenamefont {Zhou},
  \citenamefont {Gatti}, \citenamefont {Kas}, \citenamefont {Rehr},\ and\
  \citenamefont {Reining}}]{zhou_cumulant_2018}%
  \BibitemOpen
  \bibfield  {author} {\bibinfo {author} {\bibfnamefont {J.~S.}\ \bibnamefont
  {Zhou}}, \bibinfo {author} {\bibfnamefont {M.}~\bibnamefont {Gatti}},
  \bibinfo {author} {\bibfnamefont {J.~J.}\ \bibnamefont {Kas}}, \bibinfo
  {author} {\bibfnamefont {J.~J.}\ \bibnamefont {Rehr}},\ and\ \bibinfo
  {author} {\bibfnamefont {L.}~\bibnamefont {Reining}},\ }\href
  {https://doi.org/10.1103/PhysRevB.97.035137} {\bibfield  {journal} {\bibinfo
  {journal} {Physical Review B}\ }\textbf {\bibinfo {volume} {97}},\ \bibinfo
  {pages} {035137} (\bibinfo {year} {2018})},\ \bibinfo {note} {publisher:
  American Physical Society}\BibitemShut {NoStop}%
\bibitem [{\citenamefont {Caruso}\ \emph {et~al.}(2015)\citenamefont {Caruso},
  \citenamefont {Lambert},\ and\ \citenamefont {Giustino}}]{caruso_band_2015}%
  \BibitemOpen
  \bibfield  {author} {\bibinfo {author} {\bibfnamefont {F.}~\bibnamefont
  {Caruso}}, \bibinfo {author} {\bibfnamefont {H.}~\bibnamefont {Lambert}},\
  and\ \bibinfo {author} {\bibfnamefont {F.}~\bibnamefont {Giustino}},\ }\href
  {https://doi.org/10.1103/PhysRevLett.114.146404} {\bibfield  {journal}
  {\bibinfo  {journal} {Physical Review Letters}\ }\textbf {\bibinfo {volume}
  {114}},\ \bibinfo {pages} {146404} (\bibinfo {year} {2015})},\ \bibinfo
  {note} {publisher: American Physical Society}\BibitemShut {NoStop}%
\bibitem [{\citenamefont {Robinson}\ \emph
  {et~al.}(2022{\natexlab{a}})\citenamefont {Robinson}, \citenamefont {Dunn},\
  and\ \citenamefont {Reichman}}]{robinson_cumulant_2022}%
  \BibitemOpen
  \bibfield  {author} {\bibinfo {author} {\bibfnamefont {P.~J.}\ \bibnamefont
  {Robinson}}, \bibinfo {author} {\bibfnamefont {I.~S.}\ \bibnamefont {Dunn}},\
  and\ \bibinfo {author} {\bibfnamefont {D.~R.}\ \bibnamefont {Reichman}},\
  }\href {https://doi.org/10.1103/PhysRevB.105.224305} {\bibfield  {journal}
  {\bibinfo  {journal} {Physical Review B}\ }\textbf {\bibinfo {volume}
  {105}},\ \bibinfo {pages} {224305} (\bibinfo {year} {2022}{\natexlab{a}})},\
  \bibinfo {note} {publisher: American Physical Society}\BibitemShut {NoStop}%
\bibitem [{\citenamefont {Robinson}\ \emph
  {et~al.}(2022{\natexlab{b}})\citenamefont {Robinson}, \citenamefont {Dunn},\
  and\ \citenamefont {Reichman}}]{robinson_cumulant_2022-1}%
  \BibitemOpen
  \bibfield  {author} {\bibinfo {author} {\bibfnamefont {P.~J.}\ \bibnamefont
  {Robinson}}, \bibinfo {author} {\bibfnamefont {I.~S.}\ \bibnamefont {Dunn}},\
  and\ \bibinfo {author} {\bibfnamefont {D.~R.}\ \bibnamefont {Reichman}},\
  }\href {https://doi.org/10.1103/PhysRevB.105.224304} {\bibfield  {journal}
  {\bibinfo  {journal} {Physical Review B}\ }\textbf {\bibinfo {volume}
  {105}},\ \bibinfo {pages} {224304} (\bibinfo {year} {2022}{\natexlab{b}})},\
  \bibinfo {note} {publisher: American Physical Society}\BibitemShut {NoStop}%
\bibitem [{\citenamefont {Bonča}\ \emph {et~al.}(2019)\citenamefont {Bonča},
  \citenamefont {Trugman},\ and\ \citenamefont {Berciu}}]{bonca_spectral_2019}%
  \BibitemOpen
  \bibfield  {author} {\bibinfo {author} {\bibfnamefont {J.}~\bibnamefont
  {Bonča}}, \bibinfo {author} {\bibfnamefont {S.~A.}\ \bibnamefont
  {Trugman}},\ and\ \bibinfo {author} {\bibfnamefont {M.}~\bibnamefont
  {Berciu}},\ }\href {https://doi.org/10.1103/PhysRevB.100.094307} {\bibfield
  {journal} {\bibinfo  {journal} {Physical Review B}\ }\textbf {\bibinfo
  {volume} {100}},\ \bibinfo {pages} {094307} (\bibinfo {year} {2019})},\
  \bibinfo {note} {publisher: American Physical Society}\BibitemShut {NoStop}%
\bibitem [{\citenamefont {Barišić}(2004)}]{barisic_calculation_2004}%
  \BibitemOpen
  \bibfield  {author} {\bibinfo {author} {\bibfnamefont {O.~S.}\ \bibnamefont
  {Barišić}},\ }\href {https://doi.org/10.1103/PhysRevB.69.064302} {\bibfield
   {journal} {\bibinfo  {journal} {Physical Review B}\ }\textbf {\bibinfo
  {volume} {69}},\ \bibinfo {pages} {064302} (\bibinfo {year} {2004})},\
  \bibinfo {note} {publisher: American Physical Society}\BibitemShut {NoStop}%
\bibitem [{\citenamefont {De~Filippis}\ \emph {et~al.}(2005)\citenamefont
  {De~Filippis}, \citenamefont {Cataudella}, \citenamefont {Ramaglia},\ and\
  \citenamefont {Perroni}}]{de_filippis_static_2005}%
  \BibitemOpen
  \bibfield  {author} {\bibinfo {author} {\bibfnamefont {G.}~\bibnamefont
  {De~Filippis}}, \bibinfo {author} {\bibfnamefont {V.}~\bibnamefont
  {Cataudella}}, \bibinfo {author} {\bibfnamefont {V.~M.}\ \bibnamefont
  {Ramaglia}},\ and\ \bibinfo {author} {\bibfnamefont {C.~A.}\ \bibnamefont
  {Perroni}},\ }\href {https://doi.org/10.1103/PhysRevB.72.014307} {\bibfield
  {journal} {\bibinfo  {journal} {Physical Review B}\ }\textbf {\bibinfo
  {volume} {72}},\ \bibinfo {pages} {014307} (\bibinfo {year} {2005})},\
  \bibinfo {note} {publisher: American Physical Society}\BibitemShut {NoStop}%
\bibitem [{\citenamefont {Pingel}\ and\ \citenamefont
  {Neher}(2013)}]{pingel_comprehensive_2013}%
  \BibitemOpen
  \bibfield  {author} {\bibinfo {author} {\bibfnamefont {P.}~\bibnamefont
  {Pingel}}\ and\ \bibinfo {author} {\bibfnamefont {D.}~\bibnamefont {Neher}},\
  }\href {https://doi.org/10.1103/PhysRevB.87.115209} {\bibfield  {journal}
  {\bibinfo  {journal} {Physical Review B}\ }\textbf {\bibinfo {volume} {87}},\
  \bibinfo {pages} {115209} (\bibinfo {year} {2013})}\BibitemShut {NoStop}%
\bibitem [{\citenamefont {Ghosh}\ \emph {et~al.}(2018)\citenamefont {Ghosh},
  \citenamefont {Chew}, \citenamefont {Onorato}, \citenamefont {Pakhnyuk},
  \citenamefont {Luscombe}, \citenamefont {Salleo},\ and\ \citenamefont
  {Spano}}]{ghosh_spectral_2018}%
  \BibitemOpen
  \bibfield  {author} {\bibinfo {author} {\bibfnamefont {R.}~\bibnamefont
  {Ghosh}}, \bibinfo {author} {\bibfnamefont {A.~R.}\ \bibnamefont {Chew}},
  \bibinfo {author} {\bibfnamefont {J.}~\bibnamefont {Onorato}}, \bibinfo
  {author} {\bibfnamefont {V.}~\bibnamefont {Pakhnyuk}}, \bibinfo {author}
  {\bibfnamefont {C.~K.}\ \bibnamefont {Luscombe}}, \bibinfo {author}
  {\bibfnamefont {A.}~\bibnamefont {Salleo}},\ and\ \bibinfo {author}
  {\bibfnamefont {F.~C.}\ \bibnamefont {Spano}},\ }\href
  {https://doi.org/10.1021/acs.jpcc.8b03873} {\bibfield  {journal} {\bibinfo
  {journal} {The Journal of Physical Chemistry C}\ }\textbf {\bibinfo {volume}
  {122}},\ \bibinfo {pages} {18048} (\bibinfo {year} {2018})},\ \bibinfo {note}
  {publisher: American Chemical Society}\BibitemShut {NoStop}%
\bibitem [{\citenamefont {Chen}\ \emph {et~al.}(2021)\citenamefont {Chen},
  \citenamefont {Ghosh}, \citenamefont {Liu}, \citenamefont {Zozoulenko},
  \citenamefont {Fahlman},\ and\ \citenamefont
  {Braun}}]{chen_experimental_2021}%
  \BibitemOpen
  \bibfield  {author} {\bibinfo {author} {\bibfnamefont {Y.}~\bibnamefont
  {Chen}}, \bibinfo {author} {\bibfnamefont {S.}~\bibnamefont {Ghosh}},
  \bibinfo {author} {\bibfnamefont {X.}~\bibnamefont {Liu}}, \bibinfo {author}
  {\bibfnamefont {I.~V.}\ \bibnamefont {Zozoulenko}}, \bibinfo {author}
  {\bibfnamefont {M.}~\bibnamefont {Fahlman}},\ and\ \bibinfo {author}
  {\bibfnamefont {S.}~\bibnamefont {Braun}},\ }\href
  {https://doi.org/10.1021/acs.jpcc.0c08442} {\bibfield  {journal} {\bibinfo
  {journal} {The Journal of Physical Chemistry C}\ }\textbf {\bibinfo {volume}
  {125}},\ \bibinfo {pages} {937} (\bibinfo {year} {2021})}\BibitemShut
  {NoStop}%
\bibitem [{\citenamefont {Hulea}\ \emph {et~al.}(2006)\citenamefont {Hulea},
  \citenamefont {Fratini}, \citenamefont {Xie}, \citenamefont {Mulder},
  \citenamefont {Iossad}, \citenamefont {Rastelli}, \citenamefont {Ciuchi},\
  and\ \citenamefont {Morpurgo}}]{hulea_tunable_2006}%
  \BibitemOpen
  \bibfield  {author} {\bibinfo {author} {\bibfnamefont {I.~N.}\ \bibnamefont
  {Hulea}}, \bibinfo {author} {\bibfnamefont {S.}~\bibnamefont {Fratini}},
  \bibinfo {author} {\bibfnamefont {H.}~\bibnamefont {Xie}}, \bibinfo {author}
  {\bibfnamefont {C.~L.}\ \bibnamefont {Mulder}}, \bibinfo {author}
  {\bibfnamefont {N.~N.}\ \bibnamefont {Iossad}}, \bibinfo {author}
  {\bibfnamefont {G.}~\bibnamefont {Rastelli}}, \bibinfo {author}
  {\bibfnamefont {S.}~\bibnamefont {Ciuchi}},\ and\ \bibinfo {author}
  {\bibfnamefont {A.~F.}\ \bibnamefont {Morpurgo}},\ }\href
  {https://doi.org/10.1038/nmat1774} {\bibfield  {journal} {\bibinfo  {journal}
  {Nature Materials}\ }\textbf {\bibinfo {volume} {5}},\ \bibinfo {pages} {982}
  (\bibinfo {year} {2006})},\ \bibinfo {note} {number: 12 Publisher: Nature
  Publishing Group}\BibitemShut {NoStop}%
\bibitem [{\citenamefont {Ghosh}\ and\ \citenamefont
  {Spano}(2020)}]{ghosh_excitons_2020}%
  \BibitemOpen
  \bibfield  {author} {\bibinfo {author} {\bibfnamefont {R.}~\bibnamefont
  {Ghosh}}\ and\ \bibinfo {author} {\bibfnamefont {F.~C.}\ \bibnamefont
  {Spano}},\ }\href {https://doi.org/10.1021/acs.accounts.0c00349} {\bibfield
  {journal} {\bibinfo  {journal} {Accounts of Chemical Research}\ }\textbf
  {\bibinfo {volume} {53}},\ \bibinfo {pages} {2201} (\bibinfo {year}
  {2020})},\ \bibinfo {note} {publisher: American Chemical Society}\BibitemShut
  {NoStop}%
\bibitem [{\citenamefont {Acbas}\ \emph {et~al.}(2014)\citenamefont {Acbas},
  \citenamefont {Niessen}, \citenamefont {Snell},\ and\ \citenamefont
  {Markelz}}]{acbas_optical_2014}%
  \BibitemOpen
  \bibfield  {author} {\bibinfo {author} {\bibfnamefont {G.}~\bibnamefont
  {Acbas}}, \bibinfo {author} {\bibfnamefont {K.~A.}\ \bibnamefont {Niessen}},
  \bibinfo {author} {\bibfnamefont {E.~H.}\ \bibnamefont {Snell}},\ and\
  \bibinfo {author} {\bibfnamefont {A.~G.}\ \bibnamefont {Markelz}},\ }\href
  {https://doi.org/10.1038/ncomms4076} {\bibfield  {journal} {\bibinfo
  {journal} {Nature Communications}\ }\textbf {\bibinfo {volume} {5}},\
  \bibinfo {pages} {3076} (\bibinfo {year} {2014})},\ \bibinfo {note} {number:
  1 Publisher: Nature Publishing Group}\BibitemShut {NoStop}%
\bibitem [{\citenamefont {Ing}\ \emph {et~al.}(2018)\citenamefont {Ing},
  \citenamefont {El-Naggar},\ and\ \citenamefont {Hochbaum}}]{ing_going_2018}%
  \BibitemOpen
  \bibfield  {author} {\bibinfo {author} {\bibfnamefont {N.~L.}\ \bibnamefont
  {Ing}}, \bibinfo {author} {\bibfnamefont {M.~Y.}\ \bibnamefont {El-Naggar}},\
  and\ \bibinfo {author} {\bibfnamefont {A.~I.}\ \bibnamefont {Hochbaum}},\
  }\href {https://doi.org/10.1021/acs.jpcb.8b07431} {\bibfield  {journal}
  {\bibinfo  {journal} {The Journal of Physical Chemistry B}\ }\textbf
  {\bibinfo {volume} {122}},\ \bibinfo {pages} {10403} (\bibinfo {year}
  {2018})},\ \bibinfo {note} {publisher: American Chemical Society}\BibitemShut
  {NoStop}%
\bibitem [{\citenamefont {Hekstra}\ \emph {et~al.}(2016)\citenamefont
  {Hekstra}, \citenamefont {White}, \citenamefont {Socolich}, \citenamefont
  {Henning}, \citenamefont {Šrajer},\ and\ \citenamefont
  {Ranganathan}}]{hekstra_electric-field-stimulated_2016}%
  \BibitemOpen
  \bibfield  {author} {\bibinfo {author} {\bibfnamefont {D.~R.}\ \bibnamefont
  {Hekstra}}, \bibinfo {author} {\bibfnamefont {K.~I.}\ \bibnamefont {White}},
  \bibinfo {author} {\bibfnamefont {M.~A.}\ \bibnamefont {Socolich}}, \bibinfo
  {author} {\bibfnamefont {R.~W.}\ \bibnamefont {Henning}}, \bibinfo {author}
  {\bibfnamefont {V.}~\bibnamefont {Šrajer}},\ and\ \bibinfo {author}
  {\bibfnamefont {R.}~\bibnamefont {Ranganathan}},\ }\href
  {https://doi.org/10.1038/nature20571} {\bibfield  {journal} {\bibinfo
  {journal} {Nature}\ }\textbf {\bibinfo {volume} {540}},\ \bibinfo {pages}
  {400} (\bibinfo {year} {2016})},\ \bibinfo {note} {number: 7633 Publisher:
  Nature Publishing Group}\BibitemShut {NoStop}%
\bibitem [{\citenamefont {Damascelli}\ \emph {et~al.}(2003)\citenamefont
  {Damascelli}, \citenamefont {Hussain},\ and\ \citenamefont
  {Shen}}]{damascelli_angle-resolved_2003}%
  \BibitemOpen
  \bibfield  {author} {\bibinfo {author} {\bibfnamefont {A.}~\bibnamefont
  {Damascelli}}, \bibinfo {author} {\bibfnamefont {Z.}~\bibnamefont
  {Hussain}},\ and\ \bibinfo {author} {\bibfnamefont {Z.-X.}\ \bibnamefont
  {Shen}},\ }\href {https://doi.org/10.1103/RevModPhys.75.473} {\bibfield
  {journal} {\bibinfo  {journal} {Reviews of Modern Physics}\ }\textbf
  {\bibinfo {volume} {75}},\ \bibinfo {pages} {473} (\bibinfo {year} {2003})},\
  \bibinfo {note} {publisher: American Physical Society}\BibitemShut {NoStop}%
\bibitem [{\citenamefont {Baldini}\ \emph {et~al.}(2020)\citenamefont
  {Baldini}, \citenamefont {Sentef}, \citenamefont {Acharya}, \citenamefont
  {Brumme}, \citenamefont {Sheveleva}, \citenamefont {Lyzwa}, \citenamefont
  {Pomjakushina}, \citenamefont {Bernhard}, \citenamefont {van Schilfgaarde},
  \citenamefont {Carbone}, \citenamefont {Rubio},\ and\ \citenamefont
  {Weber}}]{baldini_electronphonon-driven_2020}%
  \BibitemOpen
  \bibfield  {author} {\bibinfo {author} {\bibfnamefont {E.}~\bibnamefont
  {Baldini}}, \bibinfo {author} {\bibfnamefont {M.~A.}\ \bibnamefont {Sentef}},
  \bibinfo {author} {\bibfnamefont {S.}~\bibnamefont {Acharya}}, \bibinfo
  {author} {\bibfnamefont {T.}~\bibnamefont {Brumme}}, \bibinfo {author}
  {\bibfnamefont {E.}~\bibnamefont {Sheveleva}}, \bibinfo {author}
  {\bibfnamefont {F.}~\bibnamefont {Lyzwa}}, \bibinfo {author} {\bibfnamefont
  {E.}~\bibnamefont {Pomjakushina}}, \bibinfo {author} {\bibfnamefont
  {C.}~\bibnamefont {Bernhard}}, \bibinfo {author} {\bibfnamefont
  {M.}~\bibnamefont {van Schilfgaarde}}, \bibinfo {author} {\bibfnamefont
  {F.}~\bibnamefont {Carbone}}, \bibinfo {author} {\bibfnamefont
  {A.}~\bibnamefont {Rubio}},\ and\ \bibinfo {author} {\bibfnamefont
  {C.}~\bibnamefont {Weber}},\ }\href {https://doi.org/10.1073/pnas.1919451117}
  {\bibfield  {journal} {\bibinfo  {journal} {Proceedings of the National
  Academy of Sciences}\ }\textbf {\bibinfo {volume} {117}},\ \bibinfo {pages}
  {6409} (\bibinfo {year} {2020})},\ \bibinfo {note} {publisher: Proceedings of
  the National Academy of Sciences}\BibitemShut {NoStop}%
\bibitem [{\citenamefont {Cataudella}\ \emph {et~al.}(2007)\citenamefont
  {Cataudella}, \citenamefont {De~Filippis}, \citenamefont {Mishchenko},\ and\
  \citenamefont {Nagaosa}}]{cataudella_temperature_2007}%
  \BibitemOpen
  \bibfield  {author} {\bibinfo {author} {\bibfnamefont {V.}~\bibnamefont
  {Cataudella}}, \bibinfo {author} {\bibfnamefont {G.}~\bibnamefont
  {De~Filippis}}, \bibinfo {author} {\bibfnamefont {A.~S.}\ \bibnamefont
  {Mishchenko}},\ and\ \bibinfo {author} {\bibfnamefont {N.}~\bibnamefont
  {Nagaosa}},\ }\href {https://doi.org/10.1103/PhysRevLett.99.226402}
  {\bibfield  {journal} {\bibinfo  {journal} {Physical Review Letters}\
  }\textbf {\bibinfo {volume} {99}},\ \bibinfo {pages} {226402} (\bibinfo
  {year} {2007})}\BibitemShut {NoStop}%
\bibitem [{\citenamefont {Rettig}\ \emph {et~al.}(2013)\citenamefont {Rettig},
  \citenamefont {Cortés}, \citenamefont {Jeevan}, \citenamefont {Gegenwart},
  \citenamefont {Wolf}, \citenamefont {Fink},\ and\ \citenamefont
  {Bovensiepen}}]{rettig_electronphonon_2013}%
  \BibitemOpen
  \bibfield  {author} {\bibinfo {author} {\bibfnamefont {L.}~\bibnamefont
  {Rettig}}, \bibinfo {author} {\bibfnamefont {R.}~\bibnamefont {Cortés}},
  \bibinfo {author} {\bibfnamefont {H.~S.}\ \bibnamefont {Jeevan}}, \bibinfo
  {author} {\bibfnamefont {P.}~\bibnamefont {Gegenwart}}, \bibinfo {author}
  {\bibfnamefont {T.}~\bibnamefont {Wolf}}, \bibinfo {author} {\bibfnamefont
  {J.}~\bibnamefont {Fink}},\ and\ \bibinfo {author} {\bibfnamefont
  {U.}~\bibnamefont {Bovensiepen}},\ }\href
  {https://doi.org/10.1088/1367-2630/15/8/083023} {\bibfield  {journal}
  {\bibinfo  {journal} {New Journal of Physics}\ }\textbf {\bibinfo {volume}
  {15}},\ \bibinfo {pages} {083023} (\bibinfo {year} {2013})},\ \bibinfo {note}
  {publisher: IOP Publishing}\BibitemShut {NoStop}%
\bibitem [{\citenamefont {Das}\ and\ \citenamefont
  {Sil}(1993)}]{das_small_1993}%
  \BibitemOpen
  \bibfield  {author} {\bibinfo {author} {\bibfnamefont {A.~N.}\ \bibnamefont
  {Das}}\ and\ \bibinfo {author} {\bibfnamefont {S.}~\bibnamefont {Sil}},\
  }\href {https://doi.org/10.1016/0921-4534(93)90422-M} {\bibfield  {journal}
  {\bibinfo  {journal} {Physica C: Superconductivity}\ }\textbf {\bibinfo
  {volume} {207}},\ \bibinfo {pages} {51} (\bibinfo {year} {1993})}\BibitemShut
  {NoStop}%
\bibitem [{\citenamefont {Millis}\ \emph {et~al.}(1995)\citenamefont {Millis},
  \citenamefont {Littlewood},\ and\ \citenamefont
  {Shraiman}}]{millis_double_1995}%
  \BibitemOpen
  \bibfield  {author} {\bibinfo {author} {\bibfnamefont {A.~J.}\ \bibnamefont
  {Millis}}, \bibinfo {author} {\bibfnamefont {P.~B.}\ \bibnamefont
  {Littlewood}},\ and\ \bibinfo {author} {\bibfnamefont {B.~I.}\ \bibnamefont
  {Shraiman}},\ }\href {https://doi.org/10.1103/PhysRevLett.74.5144} {\bibfield
   {journal} {\bibinfo  {journal} {Physical Review Letters}\ }\textbf {\bibinfo
  {volume} {74}},\ \bibinfo {pages} {5144} (\bibinfo {year} {1995})},\ \bibinfo
  {note} {publisher: American Physical Society}\BibitemShut {NoStop}%
\bibitem [{\citenamefont {Millis}\ \emph {et~al.}(1996)\citenamefont {Millis},
  \citenamefont {Shraiman},\ and\ \citenamefont
  {Mueller}}]{millis_dynamic_1996}%
  \BibitemOpen
  \bibfield  {author} {\bibinfo {author} {\bibfnamefont {A.}~\bibnamefont
  {Millis}}, \bibinfo {author} {\bibfnamefont {B.}~\bibnamefont {Shraiman}},\
  and\ \bibinfo {author} {\bibfnamefont {R.}~\bibnamefont {Mueller}},\ }\href
  {https://doi.org/10.1103/PhysRevLett.77.175} {\bibfield  {journal} {\bibinfo
  {journal} {Physical Review Letters}\ }\textbf {\bibinfo {volume} {77}},\
  \bibinfo {pages} {175} (\bibinfo {year} {1996})}\BibitemShut {NoStop}%
\bibitem [{\citenamefont {Zaanen}\ and\ \citenamefont
  {Littlewood}(1994)}]{zaanen_freezing_1994}%
  \BibitemOpen
  \bibfield  {author} {\bibinfo {author} {\bibfnamefont {J.}~\bibnamefont
  {Zaanen}}\ and\ \bibinfo {author} {\bibfnamefont {P.~B.}\ \bibnamefont
  {Littlewood}},\ }\href {https://doi.org/10.1103/PhysRevB.50.7222} {\bibfield
  {journal} {\bibinfo  {journal} {Physical Review B}\ }\textbf {\bibinfo
  {volume} {50}},\ \bibinfo {pages} {7222} (\bibinfo {year}
  {1994})}\BibitemShut {NoStop}%
\bibitem [{\citenamefont {Kaplan}\ and\ \citenamefont
  {Zimmerman}(1998)}]{kaplan_mechanism_1998}%
  \BibitemOpen
  \bibfield  {author} {\bibinfo {author} {\bibfnamefont {M.~D.}\ \bibnamefont
  {Kaplan}}\ and\ \bibinfo {author} {\bibfnamefont {G.~O.}\ \bibnamefont
  {Zimmerman}},\ }\href {https://doi.org/10.1016/S0022-3697(98)00217-0}
  {\bibfield  {journal} {\bibinfo  {journal} {Journal of Physics and Chemistry
  of Solids}\ }\textbf {\bibinfo {volume} {59}},\ \bibinfo {pages} {2218}
  (\bibinfo {year} {1998})}\BibitemShut {NoStop}%
\bibitem [{\citenamefont {Tang}\ \emph {et~al.}(2020)\citenamefont {Tang},
  \citenamefont {Chen}, \citenamefont {Huang}, \citenamefont {Bright},
  \citenamefont {Meng}, \citenamefont {Liu},\ and\ \citenamefont
  {Wu}}]{tang_plasmonic_2020}%
  \BibitemOpen
  \bibfield  {author} {\bibinfo {author} {\bibfnamefont {H.}~\bibnamefont
  {Tang}}, \bibinfo {author} {\bibfnamefont {C.-J.}\ \bibnamefont {Chen}},
  \bibinfo {author} {\bibfnamefont {Z.}~\bibnamefont {Huang}}, \bibinfo
  {author} {\bibfnamefont {J.}~\bibnamefont {Bright}}, \bibinfo {author}
  {\bibfnamefont {G.}~\bibnamefont {Meng}}, \bibinfo {author} {\bibfnamefont
  {R.-S.}\ \bibnamefont {Liu}},\ and\ \bibinfo {author} {\bibfnamefont
  {N.}~\bibnamefont {Wu}},\ }\href {https://doi.org/10.1063/5.0005334}
  {\bibfield  {journal} {\bibinfo  {journal} {The Journal of Chemical Physics}\
  }\textbf {\bibinfo {volume} {152}},\ \bibinfo {pages} {220901} (\bibinfo
  {year} {2020})},\ \bibinfo {note} {publisher: American Institute of
  Physics}\BibitemShut {NoStop}%
\bibitem [{\citenamefont {Cushing}\ and\ \citenamefont
  {Wu}(2016)}]{cushing_progress_2016}%
  \BibitemOpen
  \bibfield  {author} {\bibinfo {author} {\bibfnamefont {S.~K.}\ \bibnamefont
  {Cushing}}\ and\ \bibinfo {author} {\bibfnamefont {N.}~\bibnamefont {Wu}},\
  }\href {https://doi.org/10.1021/acs.jpclett.5b02393} {\bibfield  {journal}
  {\bibinfo  {journal} {The Journal of Physical Chemistry Letters}\ }\textbf
  {\bibinfo {volume} {7}},\ \bibinfo {pages} {666} (\bibinfo {year} {2016})},\
  \bibinfo {note} {publisher: American Chemical Society}\BibitemShut {NoStop}%
\bibitem [{\citenamefont {Ma}\ \emph {et~al.}(2016)\citenamefont {Ma},
  \citenamefont {Dai}, \citenamefont {Yu},\ and\ \citenamefont
  {Huang}}]{ma_energy_2016}%
  \BibitemOpen
  \bibfield  {author} {\bibinfo {author} {\bibfnamefont {X.-C.}\ \bibnamefont
  {Ma}}, \bibinfo {author} {\bibfnamefont {Y.}~\bibnamefont {Dai}}, \bibinfo
  {author} {\bibfnamefont {L.}~\bibnamefont {Yu}},\ and\ \bibinfo {author}
  {\bibfnamefont {B.-B.}\ \bibnamefont {Huang}},\ }\href
  {https://doi.org/10.1038/lsa.2016.17} {\bibfield  {journal} {\bibinfo
  {journal} {Light: Science \& Applications}\ }\textbf {\bibinfo {volume}
  {5}},\ \bibinfo {pages} {e16017} (\bibinfo {year} {2016})},\ \bibinfo {note}
  {number: 2 Publisher: Nature Publishing Group}\BibitemShut {NoStop}%
\bibitem [{\citenamefont {Melnick}\ and\ \citenamefont
  {Kaviany}(2016{\natexlab{a}})}]{melnick_phonovoltaic_2016}%
  \BibitemOpen
  \bibfield  {author} {\bibinfo {author} {\bibfnamefont {C.}~\bibnamefont
  {Melnick}}\ and\ \bibinfo {author} {\bibfnamefont {M.}~\bibnamefont
  {Kaviany}},\ }\href {https://doi.org/10.1103/PhysRevB.93.094302} {\bibfield
  {journal} {\bibinfo  {journal} {Physical Review B}\ }\textbf {\bibinfo
  {volume} {93}},\ \bibinfo {pages} {094302} (\bibinfo {year}
  {2016}{\natexlab{a}})},\ \bibinfo {note} {publisher: American Physical
  Society}\BibitemShut {NoStop}%
\bibitem [{\citenamefont {Melnick}\ and\ \citenamefont
  {Kaviany}(2016{\natexlab{b}})}]{melnick_phonovoltaic_2016-2}%
  \BibitemOpen
  \bibfield  {author} {\bibinfo {author} {\bibfnamefont {C.}~\bibnamefont
  {Melnick}}\ and\ \bibinfo {author} {\bibfnamefont {M.}~\bibnamefont
  {Kaviany}},\ }\href {https://doi.org/10.1103/PhysRevB.94.245412} {\bibfield
  {journal} {\bibinfo  {journal} {Physical Review B}\ }\textbf {\bibinfo
  {volume} {94}},\ \bibinfo {pages} {245412} (\bibinfo {year}
  {2016}{\natexlab{b}})},\ \bibinfo {note} {publisher: American Physical
  Society}\BibitemShut {NoStop}%
\bibitem [{\citenamefont {Riley}\ \emph {et~al.}(2018)\citenamefont {Riley},
  \citenamefont {Caruso}, \citenamefont {Verdi}, \citenamefont {Duffy},
  \citenamefont {Watson}, \citenamefont {Bawden}, \citenamefont {Volckaert},
  \citenamefont {van~der Laan}, \citenamefont {Hesjedal}, \citenamefont
  {Hoesch}, \citenamefont {Giustino},\ and\ \citenamefont
  {King}}]{riley_crossover_2018}%
  \BibitemOpen
  \bibfield  {author} {\bibinfo {author} {\bibfnamefont {J.~M.}\ \bibnamefont
  {Riley}}, \bibinfo {author} {\bibfnamefont {F.}~\bibnamefont {Caruso}},
  \bibinfo {author} {\bibfnamefont {C.}~\bibnamefont {Verdi}}, \bibinfo
  {author} {\bibfnamefont {L.~B.}\ \bibnamefont {Duffy}}, \bibinfo {author}
  {\bibfnamefont {M.~D.}\ \bibnamefont {Watson}}, \bibinfo {author}
  {\bibfnamefont {L.}~\bibnamefont {Bawden}}, \bibinfo {author} {\bibfnamefont
  {K.}~\bibnamefont {Volckaert}}, \bibinfo {author} {\bibfnamefont
  {G.}~\bibnamefont {van~der Laan}}, \bibinfo {author} {\bibfnamefont
  {T.}~\bibnamefont {Hesjedal}}, \bibinfo {author} {\bibfnamefont
  {M.}~\bibnamefont {Hoesch}}, \bibinfo {author} {\bibfnamefont
  {F.}~\bibnamefont {Giustino}},\ and\ \bibinfo {author} {\bibfnamefont
  {P.~D.~C.}\ \bibnamefont {King}},\ }\href
  {https://doi.org/10.1038/s41467-018-04749-w} {\bibfield  {journal} {\bibinfo
  {journal} {Nature Communications}\ }\textbf {\bibinfo {volume} {9}},\
  \bibinfo {pages} {2305} (\bibinfo {year} {2018})},\ \bibinfo {note} {number:
  1 Publisher: Nature Publishing Group}\BibitemShut {NoStop}%
\bibitem [{\citenamefont {Holstein}(1959)}]{holstein_studies_1959}%
  \BibitemOpen
  \bibfield  {author} {\bibinfo {author} {\bibfnamefont {T.}~\bibnamefont
  {Holstein}},\ }\href {https://doi.org/10.1016/0003-4916(59)90002-8}
  {\bibfield  {journal} {\bibinfo  {journal} {Annals of Physics}\ }\textbf
  {\bibinfo {volume} {8}},\ \bibinfo {pages} {325} (\bibinfo {year}
  {1959})}\BibitemShut {NoStop}%
\bibitem [{\citenamefont {Mahan}(2000)}]{mahan_many-particle_2000}%
  \BibitemOpen
  \bibfield  {author} {\bibinfo {author} {\bibfnamefont {G.~D.}\ \bibnamefont
  {Mahan}},\ }\href {https://doi.org/10.1007/978-1-4757-5714-9} {\emph
  {\bibinfo {title} {Many-{Particle} {Physics}}}}\ (\bibinfo  {publisher}
  {Springer US},\ \bibinfo {address} {Boston, MA},\ \bibinfo {year}
  {2000})\BibitemShut {NoStop}%
\bibitem [{\citenamefont {Luchner}\ and\ \citenamefont
  {Micklitz}(1979)}]{luchner_phonon_1979}%
  \BibitemOpen
  \bibfield  {author} {\bibinfo {author} {\bibfnamefont {K.}~\bibnamefont
  {Luchner}}\ and\ \bibinfo {author} {\bibfnamefont {H.}~\bibnamefont
  {Micklitz}},\ }\href {https://doi.org/10.1016/0022-2313(79)90256-4}
  {\bibfield  {journal} {\bibinfo  {journal} {Journal of Luminescence}\
  }\textbf {\bibinfo {volume} {18-19}},\ \bibinfo {pages} {882} (\bibinfo
  {year} {1979})}\BibitemShut {NoStop}%
\bibitem [{\citenamefont {Marchand}\ and\ \citenamefont
  {Berciu}(2013)}]{marchand_effect_2013}%
  \BibitemOpen
  \bibfield  {author} {\bibinfo {author} {\bibfnamefont {D.~J.~J.}\
  \bibnamefont {Marchand}}\ and\ \bibinfo {author} {\bibfnamefont
  {M.}~\bibnamefont {Berciu}},\ }\href
  {https://doi.org/10.1103/PhysRevB.88.060301} {\bibfield  {journal} {\bibinfo
  {journal} {Physical Review B}\ }\textbf {\bibinfo {volume} {88}},\ \bibinfo
  {pages} {060301} (\bibinfo {year} {2013})},\ \bibinfo {note} {publisher:
  American Physical Society}\BibitemShut {NoStop}%
\bibitem [{\citenamefont {Bonča}\ and\ \citenamefont
  {Trugman}(2021)}]{bonca_dynamic_2021}%
  \BibitemOpen
  \bibfield  {author} {\bibinfo {author} {\bibfnamefont {J.}~\bibnamefont
  {Bonča}}\ and\ \bibinfo {author} {\bibfnamefont {S.~A.}\ \bibnamefont
  {Trugman}},\ }\href {https://doi.org/10.1103/PhysRevB.103.054304} {\bibfield
  {journal} {\bibinfo  {journal} {Physical Review B}\ }\textbf {\bibinfo
  {volume} {103}},\ \bibinfo {pages} {054304} (\bibinfo {year} {2021})},\
  \bibinfo {note} {publisher: American Physical Society}\BibitemShut {NoStop}%
\bibitem [{\citenamefont {Migdal}(1958)}]{migdal_interaction_1958}%
  \BibitemOpen
  \bibfield  {author} {\bibinfo {author} {\bibfnamefont {A.~B.}\ \bibnamefont
  {Migdal}},\ }\href {https://www.osti.gov/biblio/4296936} {\bibfield
  {journal} {\bibinfo  {journal} {Zhur. Eksptl'. i Teoret. Fiz.}\ }\textbf
  {\bibinfo {volume} {Vol: 34}} (\bibinfo {year} {1958})},\ \bibinfo {note}
  {institution: Moscow Inst. of Engineering Physics}\BibitemShut {NoStop}%
\end{thebibliography}%

\end{document}